\documentclass[preprint,prl,onecolumn,showpacs,amsmath,amssymb,superscriptaddress,preprintnumbers]{revtex4-1}
\usepackage[english]{babel}
\usepackage{amsmath,amssymb,amsfonts} 	
\usepackage{graphicx}
\usepackage[utf8]{inputenc}
\usepackage{bm}
\usepackage{xcolor}
\usepackage[normalem]{ulem}
\usepackage{SIunits}
\usepackage[colorlinks=true,linkcolor=blue,citecolor=blue,urlcolor=blue]{hyperref}
\usepackage{url}
\usepackage{textcomp}
\usepackage{lineno,hyperref}
\usepackage{rotating}
\usepackage{color} 
\definecolor{red}{rgb}{1.,0.,0.}
\usepackage{soul}
\usepackage{setspace}
\usepackage{braket}
\usepackage{caption}
\usepackage{subfig}

\newcommand{\noi}{\noindent}
\newcommand{\mb}{\mathbf}

\newcommand\THEOSMARVEL{Theory and Simulation of Materials (THEOS) and National Centre for Computational Design and Discovery of Novel Materials (MARVEL), {\'E}cole Polytechnique F{\'e}d{\'e}rale de Lausanne, 1015 Lausanne, Switzerland}
\newcommand\Louvain{Institute of Condensed Matter and Nanosciences (IMCN), Université catholique de Louivain, 1348 Louvain-la-Neuve, Belgium}

\begin{document}

\title{Photorealistic modelling of metals from first principles}
\author{Gianluca Prandini}
\affiliation{\THEOSMARVEL}
\author{Gian-Marco Rignanese}
\affiliation{\Louvain}
\author{Nicola Marzari}
\email{nicola.marzari@epfl.ch}
\affiliation{\THEOSMARVEL}

\begin{abstract}
\noi
The colours of metals have attracted the attention of humanity since ancient times, and coloured metals, in particular gold compounds, have been employed for tools and objects symbolizing the aesthetics of power. 
In this work we develop a comprehensive framework to obtain the reflectivity and colour of metals, and show that the trends in optical properties and the colours can be predicted by straightforward first-principles techniques based on standard approximations. We apply this to predict reflectivity and colour of several elemental metals and of different types of metallic compounds (intermetallics, solid solutions and heterogeneous alloys), considering mainly binary alloys based on noble metals. We validate the numerical approach through an extensive comparison with experimental data and the photorealistic rendering of known coloured metals.
\end{abstract}

\maketitle


\section{Introduction}
\noindent
Optical properties of metals are important for novel technological applications where the optical response needs to be engineered for specific purposes, such as for plasmonic devices (e.g. in spectrally-selective coatings~\cite{Bilokur2017, Guo2014}), for optoelectronics devices (e.g. in ultra-thin films for transparent conductive electrodes~\cite{Yun2017, Ren2015}), and also for microscopy and optical data storage~\cite{Hatwar1997}.
Also, the colours of metals (which are related to the optical properties within the visible range of the electromagnetic spectrum) play a significant role in the jewellery industry, decoration and dentistry.
For these applications, the most used materials are metallic alloys based on gold or other coinage or precious metals, such as silver, copper, palladium and platinum.  
In particular, gold and copper are the few elemental metals that show a characteristic colour, due to the presence of a drop in the reflectivity curve inside the visible range; reflectivities of nearly all other metals are instead generally high and flat for all visible frequencies, making them appear shiny and silvery white.
Moreover, gold alloys and intermetallics are known to show a broad spectrum of colours (yellow, red, purple, to name a few), which can be tuned by varying the alloying elements and concentrations in the material~\cite{Cretu1999}. 
Since in the jewellery industry there is the demand, due to market and fashion trends, for new precious-metal alloys with specific colours, it is also of great interest there the search and identification of novel alloys with novel optical properties. \\
Generally speaking, the common route followed by researchers and manufacturers in order to identify any type of novel materials is through trial-and-error experiments which, however, have the drawback to be time-consuming and, if dealing with precious-metal-based systems, expensive. 
In order to streamline this process, an alternative route that can help in guiding the search for new promising candidate systems is computational modelling, so that the physical properties under investigation are assessed through computer simulations rather than by real experiments. 
Here, we show and discuss how it is possible to perform photorealistic simulations of metals by means of first-principles methods and, as a consequence, predict the colour of novel metallic alloys. 
Previously published studies about first-principles simulation of optical properties of both elemental metals and alloys already point towards  
the feasibility of this approach. Indeed, in 1988 Maksimov \textit{et al}.~\cite{Maksimov1988} computed the optical properties of several elemental metals whereas, more recently, Werner \textit{et al}.~\cite{Werner2009} performed a similar study on 17 elemental metals; both studies found qualitative agreement with experimental results.
For compounds, on the other hand, Blaber \textit{et al}.~\cite{Blaber2009} calculated the optical properties of several intermetallic compounds, with a particular focus on alkali-noble intermetallics, for new possible candidates as plasmonic materials while, in another work, Keast \textit{et al}.~\cite{Keast2014} computed the density of states and  dielectric function of gold intermetallics compounds and gold binary alloys.
Regarding the simulation of specific coloured intermetallic compounds, the reflectivity and colour of the three well-known coloured gold intermetallics AuAl$_2$, AuGa$_2$ and AuIn$_2$ was first computed in Ref.~\cite{Keast2011} and, afterwards, Keast \textit{et al}.~\cite{Keast2013} studied the influence of alloying concentrations on the reflectivity and colour of the intermetallic AuAl$_2$ by considering the ternary compounds having the Au$_{1-x}$Pt$_x$Al$_2$ composition, with $x=0.0, 0.5, 0.75, 1.0$; equivalent computational results were obtained independently by Kecik~\cite{Kecik2013}. Calculated and experimental~\cite{Vishnubhatla1967, Furrer2014} reflectivity curves and colours for these compounds showed good agreement and the trends in colour as a function of the composition were well reproduced.
In addition, the effect of disorder on the optical properties of Au$_{0.5}$Cu$_{0.5}$ was studied by comparing the dielectric function of the random solid solution, simulated using the supercell approach, with that of the ordered intermetallic compound~\cite{DeSilva2015}, and the main spectral differences between the two different types of compounds were captured by the simulations. \\
In this work, first we establish a general computational approach that can be used for the photorealistic simulation of metals, showing how the reflectivity and colour of metallic crystals can be estimated by means of first-principles techniques. 
We then demonstrate through a systematic study on elemental metals and extensive comparisons with experimental data that the theoretical and numerical approximations adopted are able to reproduce the correct behaviour of the reflectivity curves and to capture the main differences in optical properties across the periodic table.
Finally, we perform a similar study on metal alloys by considering different types of compounds, i.e. ordered intermetallics, disordered solid solutions and heterogeneous alloys. 
In particular, we show through a comparison with experimental results that, if the appropriate methods are used for the simulation of the different types of compounds, (i) the simulated colours of known coloured intermetallics are in qualitative and most often in quantitative agreement with experiments and that (ii) one can reproduce the main colour trends in noble-metal-based binary alloys.

\section{Results}
\noindent

\subsection{Computational approach}

\begin{figure}[!hbtp]
\centering
\includegraphics[width=7cm]{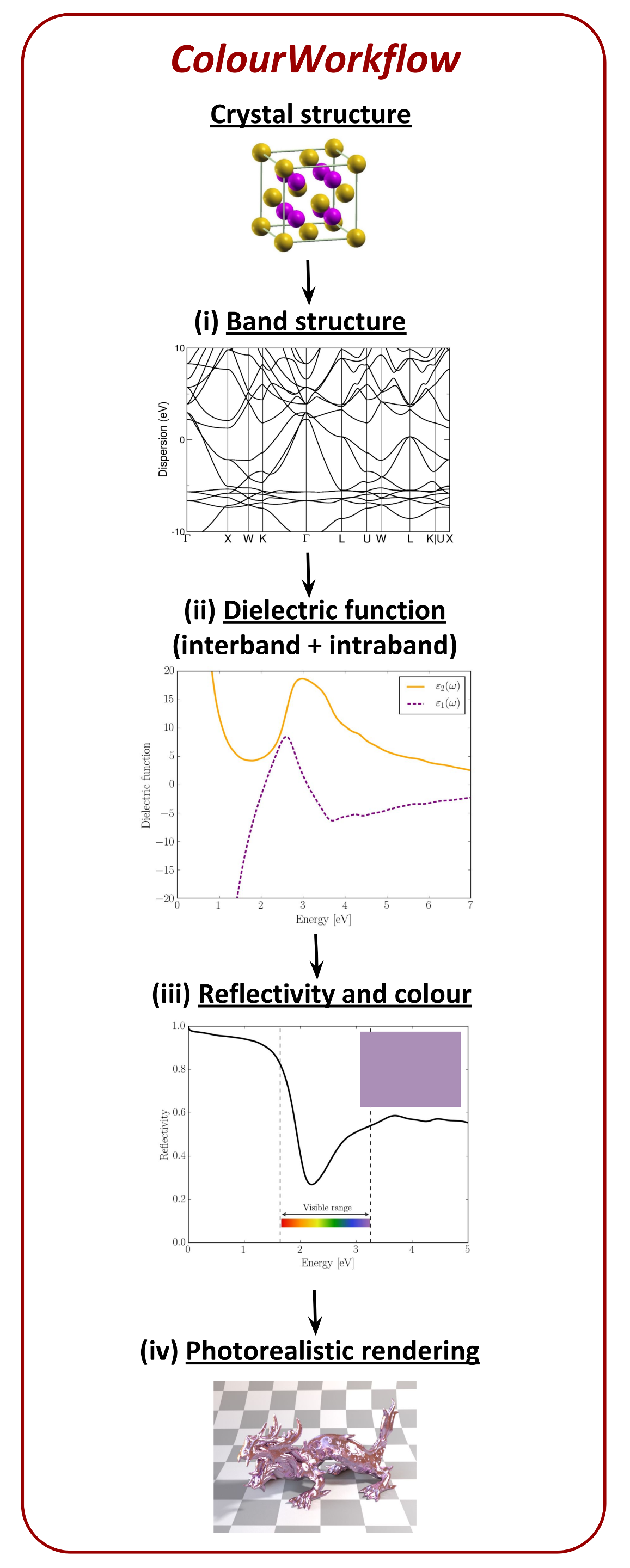}
\caption{Schematic representation of the workflow, named ColourWorkflow, designed to simulate the reflectivity and colour of a metallic material giving as input its crystal structure.  
}
\label{fig:colourworkflow}
\end{figure}

\noindent
The computational workflow that allows one to obtain the reflectivity and colour of a given metal from an initial crystal structure, schematically depicted in Fig.~\ref{fig:colourworkflow}, can be divided into four main computational steps: (i) evaluation of the electronic structure, (ii) calculation of the dielectric function, (iii) calculation of the reflectivity and colour and (iv) photorealistic rendering of the material. \\
The quantity we consider for the first-principles simulation of optical properties is the complex, wavevector- and frequency-dependent, dielectric function $\varepsilon(\mb{q},\omega) = \varepsilon_1(\mb{q},\omega) + i \varepsilon_2(\mb{q},\omega)$. 
In fact, the knowledge of the dielectric function gives then access to all the optical constants measurable by optical experiments, such as absorption coefficient and reflectivity. \\
Throughout this work the electronic structure is computed using density-functional theory (DFT)~\cite{Kohn1964dft} within the generalized gradient approximation (GGA) and relying on the PBE exchange-correlation functional~\cite{Perdew1996}. We emphasize here that the electronic structure could alternatively be obtained with more accurate techniques; for example, the accuracy of the band structures could be improved by computing quasi-particle corrections on top of PBE results (typically at the $GW$ level~\cite{Hybertsen1986,Onida2002,Reining2017}), albeit at a largely increased computational cost.
So, while in the present work we just rely on the Kohn-Sham (KS) PBE bands~\cite{Kohn1965}, the use of conceptually and quantitatively correct $GW$ bands would take place seamlessly inside this workflow.
Subsequently, we calculate the dielectric function within the independent particle approximation (IPA), which amounts to neglecting (i) effects related to electron-hole interactions (excitonic effects), since these are effectively screened by the conduction electrons and (ii) effects related to the rapidly varying microscopic electric fields inside the material (local-field effects) since these are typically small in homogeneous systems such as bulk metals~\cite{Marini2001}. 
In the optical regime the momentum $\mb{q}$ transferred by the photon is negligible so that we can consider the optical limit, $\mb{q} \to \mb{0}$, of the expression for the IPA dielectric function $\varepsilon(\mb{q},\omega)$.
In general, the dielectric function still depends on the direction $\hat{\mb{q}} = \mb{q}/|\mb{q}|$ of the perturbing electric field and only for crystals with cubic symmetry it is the same in every direction. 
In the optical limit it is convenient to divide the evaluation of the IPA dielectric function of metals into two separate contributions, an intraband Drude-like term $\varepsilon^{\text{intra}}(\hat{\mb{q}},\omega)$ due to the conduction electrons at the Fermi surface and an interband term $\varepsilon^{\text{inter}}(\hat{\mb{q}},\omega)$ due to vertical transitions between occupied and unoccupied bands, so that $\varepsilon(\hat{\mb{q}},\omega) = \varepsilon^{\text{inter}}(\hat{\mb{q}},\omega) + \varepsilon^{\text{intra}}(\hat{\mb{q}},\omega)$. 
Using the solutions of the one-particle Schr\"{o}dinger equation for periodic systems, $H^{\text{KS}}\ket{\psi_{n\mb{k}}}=E_{n\mb{k} }\ket{\psi_{n\mb{k}}}$ (where $H^{\text{KS}}$ is the KS Hamiltonian from DFT), the explicit expression of the IPA dielectric function can be written as~\cite{Wooten1972, Marini2001, Harl2008}
\begin{align} \label{eq: eps_IP_inter}
\varepsilon^{\text{inter}}(\hat{\mb{q}},\omega) &= 1  - \frac{4\pi}{V} \sum_{\mb{k}} \sum_{\substack{n, n' \\ n \neq n'}} 
\frac{| \bra{\psi_{n'\mb{k}}}  \hat{\mb{q}} \cdot \mb{v} \ket{\psi_{n\mb{k}}}  |^2 }{(E_{n'\mb{k}} -  E_{n\mb{k}})^2}
\frac{ f_{n\mb{k}} - f_{n'\mb{k}}   }{ \omega - ( E_{n'\mb{k}} -  E_{n\mb{k} } )  
 +  i\eta   }, \\ \label{eq: eps_IP_intra_dissipation}
\varepsilon^{\text{intra}}(\hat{\mb{q}},\omega) &=  - \frac{\omega^2_{\text{D}}(\hat{\mb{q}})}{\omega(\omega + i\gamma)}, 
\end{align}
\noi
where we have defined the IPA Drude plasma frequency as
\begin{equation} \label{eq: drude_plasma_freq}
\omega^2_{\text{D}}(\hat{\mb{q}})  = \frac{4\pi}{V} \sum_{\mb{k}} \sum_{n} | \bra{\psi_{n\mb{k}}}  \hat{\mb{q}} \cdot \mb{v} \ket{\psi_{n\mb{k}}}  |^2  \left(  -\frac{\partial f_{n\mb{k}}}{\partial E_{n\mb{k}} } \right).
\end{equation}
\noi
In the expressions above $\mb{v}=  -i \, [\mb{r},H^{\text{KS}}]$ is the velocity operator, $f_{n\mb{k}}$ is the Fermi-Dirac occupation function of the KS Bloch state $\ket{\psi_{n\mb{k}}}$ identified by band index $n$ and wavevector $\mb{k}$ within the Brillouin zone (BZ) and $V$ is the volume of the crystal.
Instead $\eta$ is an infinitesimal broadening introduced to perform the adiabatic switch of the perturbation within linear-response theory and, in practical calculations, 
it is used as an empirical broadening which accounts for scattering processes, always present in real materials, and/or for finite experimental resolution. Similarly, $\gamma$ is an empirical broadening representing dissipation effects of the conduction electrons (see the Methods section for more details on the parameters effectively used in the simulations). A more extensive discussion on the first-principles theory of optical properties and the derivation of the expression of the IPA dielectric function  in the optical limit (Eq.~\ref{eq: eps_IP_inter}, Eq.~\ref{eq: eps_IP_intra_dissipation} and Eq.~\ref{eq: drude_plasma_freq}) can be found in Ref.~\cite{Prandini2019}. \\
As the most typical experimental situation is to have polycrystalline materials in which grains have random orientations, in the following we always deal with the dielectric function averaged over the three Cartesian directions
\begin{equation} \label{eq: eps_average_cartesian}
\varepsilon(\omega) = \frac{ \varepsilon(\hat{\mb{x}},\omega) + \varepsilon(\hat{\mb{y}},\omega) + \varepsilon(\hat{\mb{z}},\omega) }{3},
\end{equation}
\noi
so that we can drop the dependence on the direction $\hat{\mb{q}}$. Similarly we also define a corresponding average IPA Drude plasma frequency as $\omega^2_{\text{D}} = [ \omega^2_{\text{D}}(\hat{\mb{x}}) + \omega^2_{\text{D}}(\hat{\mb{y}}) + \omega^2_{\text{D}}(\hat{\mb{z}}) ]/3$. \\
In order to compute the reflectivity from the knowledge of the dielectric function we first introduce the refractive index $n(\omega)$ and the extinction coefficient $k(\omega)$ that are defined from the equation $[n(\omega) + ik(\omega)]^2 =  \varepsilon(\omega)$. 
The reflectivity at normal incidence and assuming a vacuum-material interface is then simply linked to $n(\omega)$ and $k(\omega)$ through the Fresnel equations of classical electromagnetism (see for example Ref.~\cite{Griffiths2007}):
\begin{equation} \label{eq: reflectivity_refractive}
R(\omega) = \frac{\left[ n(\omega) -1  \right]^2 + k(\omega)^2  }{\left[ n(\omega) + 1  \right]^2 + k(\omega)^2}.
\end{equation}

\noi
Eventually, we relate the reflectivity of a material to its perceived colour using the standard colour spaces introduced by the \textit{Commission Internationale de l'Eclairage} (CIE)~\cite{cie_webpage}  for quantitative measures of colour.
For this purpose trichromatic theory gives the rigorous mathematical framework that permits to estimate the colour of an opaque material (e.g. a metal) by knowing its reflectivity $R(\lambda)$ for all the wavelengths $\lambda$ in the visible range (i.e. in the range [380, 780] nm), and to condense this information into three numbers, i.e. the colour coordinates~\cite{Schanda2007}. 
In particular, according to the \textit{CIE 1931 standard colorimetric observer}, the tristimulus values ($X$, $Y$, $Z$) which define the CIE-$XYZ$ colour space completely describe a colour stimulus and are given by the following integrals over the visible range
\begin{eqnarray} 
X & = & k\int\limits_{380 \, \text{nm}}^{780 \, \text{nm}} d\lambda \, \bar{\text{x}}(\lambda)R(\lambda)S(\lambda) , \\
Y & = & k\int\limits_{380 \, \text{nm}}^{780 \, \text{nm}} d\lambda \, \bar{\text{y}}(\lambda)R(\lambda)S(\lambda) , \\
Z & = & k\int\limits_{380 \, \text{nm}}^{780 \, \text{nm}} d\lambda \, \bar{\text{z}}(\lambda)R(\lambda)S(\lambda) ,
\end{eqnarray}
\noi
where $\bar{\text{x}}(\lambda)$, $\bar{\text{y}}(\lambda)$ and $\bar{\text{z}}(\lambda)$ are the three so-called colour-matching functions and describe the chromatic response of the observer, being related to the sensitivity of the three different colour-sensitive photoreceptors present in the human eye.
$S(\lambda)$ is instead the spectral power distribution of one of the standard CIE illuminant (throughout this work the D65 illuminant is used which corresponds to average daylight) while the constant $k$ is chosen so that $Y = 100$ for objects for which $R(\lambda) = 1$ for all visible wavelengths.\\
In practice, it is more convenient to work within the CIELAB colour space rather than in the CIE-$XYZ$ colour space, which is defined by three coordinates ($L^*$, $a^*$, $b^*$) that are easily computed from the knowledge of the tristimulus values ($X$, $Y$, $Z$) through a coordinate transformation~\cite{Schanda2007}.
Indeed, since the CIELAB colour space is nearly uniform, euclidean distances can be used to approximately represent the perceived magnitude of colour differences between two objects in the same external conditions. Therefore, if ($L_1^*$, $a_1^*$, $b_1^*$)  and ($L_2^*$, $a_2^*$, $b_2^*$) are the CIELAB coordinates of two objects, their colour difference is simply given by 
\begin{equation}
\label{eq: delta_e}
\Delta E = \sqrt{(L_1^* - L_2^*)^2 + (a_1^*-a_2^*)+(b_1^*-b_2^*)^2}.
\end{equation}
\noi
Typically, a difference $\Delta E > 1-2$ can be perceived by the human eye. 
In addition, we use photorealistic rendering, which is based on the solution of the light-transport equation~\cite{Kajiya1986}, to simulate the actual appearance of an object 
made of a material with specified optical constants in the visible range within a realistic 3D scene.

\noi
Our goal is to apply the computational approach described above to study metals in their crystalline form.
From the point of view of first-principles calculations, elemental crystals are the easiest and most computationally efficient systems to simulate since they are periodic and their primitive cell, which typically consists of only a few atoms, can simply be taken as the simulation cell.
For multi-component systems instead (in this work we focus on binary alloys), we distinguish different types of compounds according to their atomic configuration and microstructure. 
In particular, we consider the following three limiting cases: (i) perfectly ordered phases, i.e. pure intermetallic compounds, (ii) perfectly disordered phases, i.e. pure solid solutions, and (iii) heterogeneous alloys, i.e. alloys consisting of a mixture of two different phases.
Since we are exclusively interested in the study of the intrinsic bulk colours of metals, we neglect the influence on the optical properties of defects (e.g. vacancies, dislocations, etc.) and any type of surface effects.  \\
We use different simulation methods in order to properly model the reflectivity and colour of these three different types of compounds, as summarized in Table~\ref{tab:alloy-type_simulations}.
As for the case of elemental crystals, pure intermetallic compounds are periodic systems and are simply simulated in their primitive cell. 
On the other hand, we use the supercell approach, based on the use of special quasi-random structures (SQS)~\cite{Zunger1990, Wei1990}, 
to take into account effects related to disorder in the simulation of the optical properties of solid solutions (see Supplementary Discussion 1 for a comparison between the SQS supercell approach and the virtual-crystal approximation).
Instead, for heterogeneous alloys made of two phases $\alpha$ and $\beta$, we use the Bruggeman model~\cite{Niklasson1981} to estimate the optical properties of the alloy. 
Within the Bruggeman model the dielectric function of the mixture, that we indicate as $\varepsilon_{\text{Br}}(\omega)$, is given by the following expression
\begin{equation} \label{eq: bruggeman_model}
(1-x_{\beta}) \frac{ \varepsilon_{\alpha}(\omega) - \varepsilon_{\text{Br}}(\omega) }{\varepsilon_{\alpha}(\omega) + 2\varepsilon_{\text{Br}}(\omega)} + x_{\beta} \frac{ \varepsilon_{\beta}(\omega) - \varepsilon_{\text{Br}}(\omega) }{\varepsilon_{\beta}(\omega) + 2\varepsilon_{\text{Br}}(\omega)} = 0,
\end{equation}
\noi
in terms of the dielectric functions $\varepsilon_{\alpha}(\omega)$ and $\varepsilon_{\beta}(\omega)$ of the single phases, and where $x_{\alpha}$ and $x_{\beta}$ (with $x_{\alpha} + x_{\beta} = 1$) are the fractions of the two phases present in the material.
The dielectric function of the single phases can be obtained with the methods of Table~\ref{tab:alloy-type_simulations} for intermetallic compounds and solid solutions. \\
In the following, we apply and validate the computational approach described here, and discuss its limitations, on several elemental metals and binary compounds.

\subsection{Elemental metals}
\begin{figure}[!hbtp]
\centering
\subfloat{\includegraphics[scale=0.37]{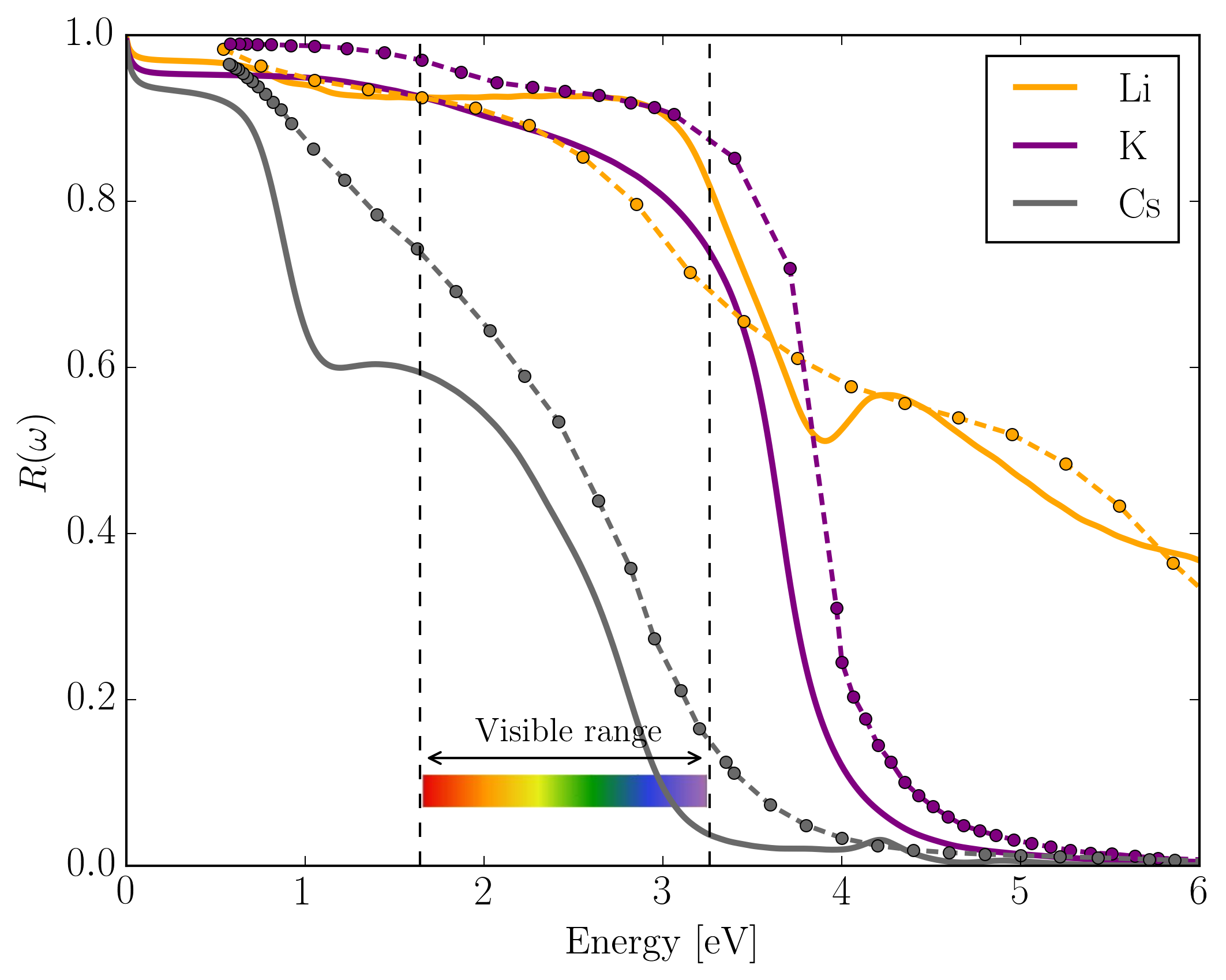}}
\subfloat{\includegraphics[scale=0.37]{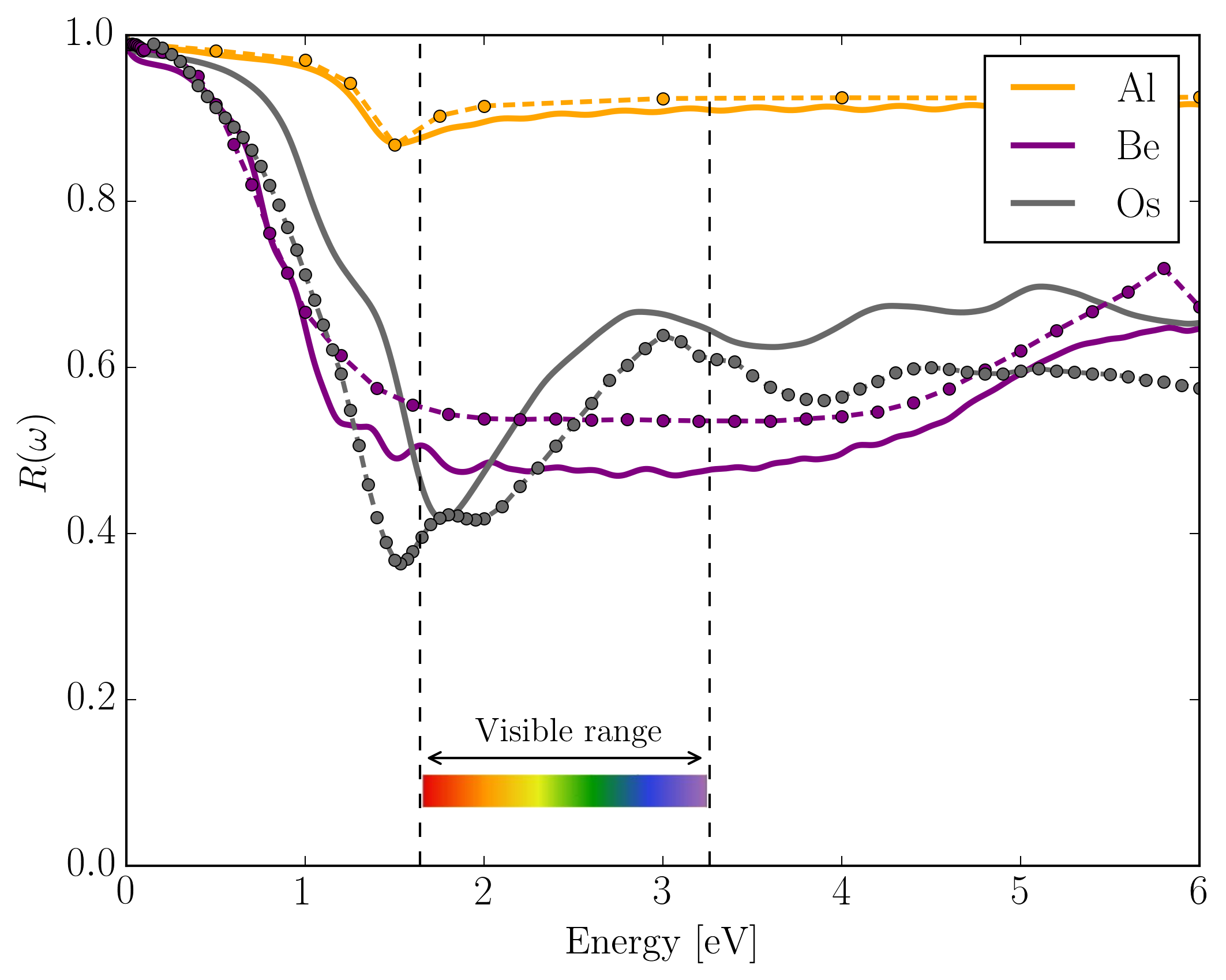}}

\subfloat{\includegraphics[scale=0.37]{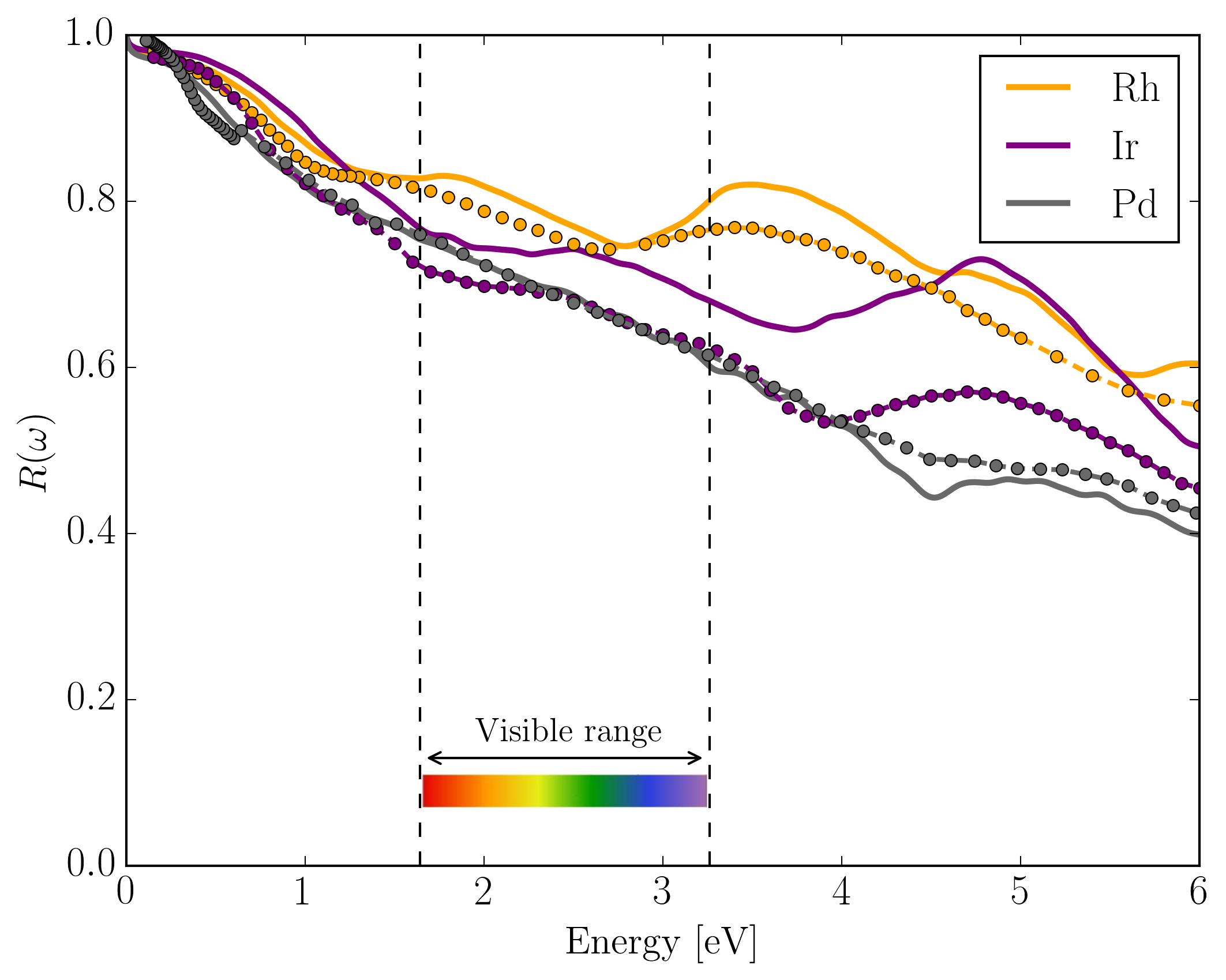}}
\subfloat{\includegraphics[scale=0.37]{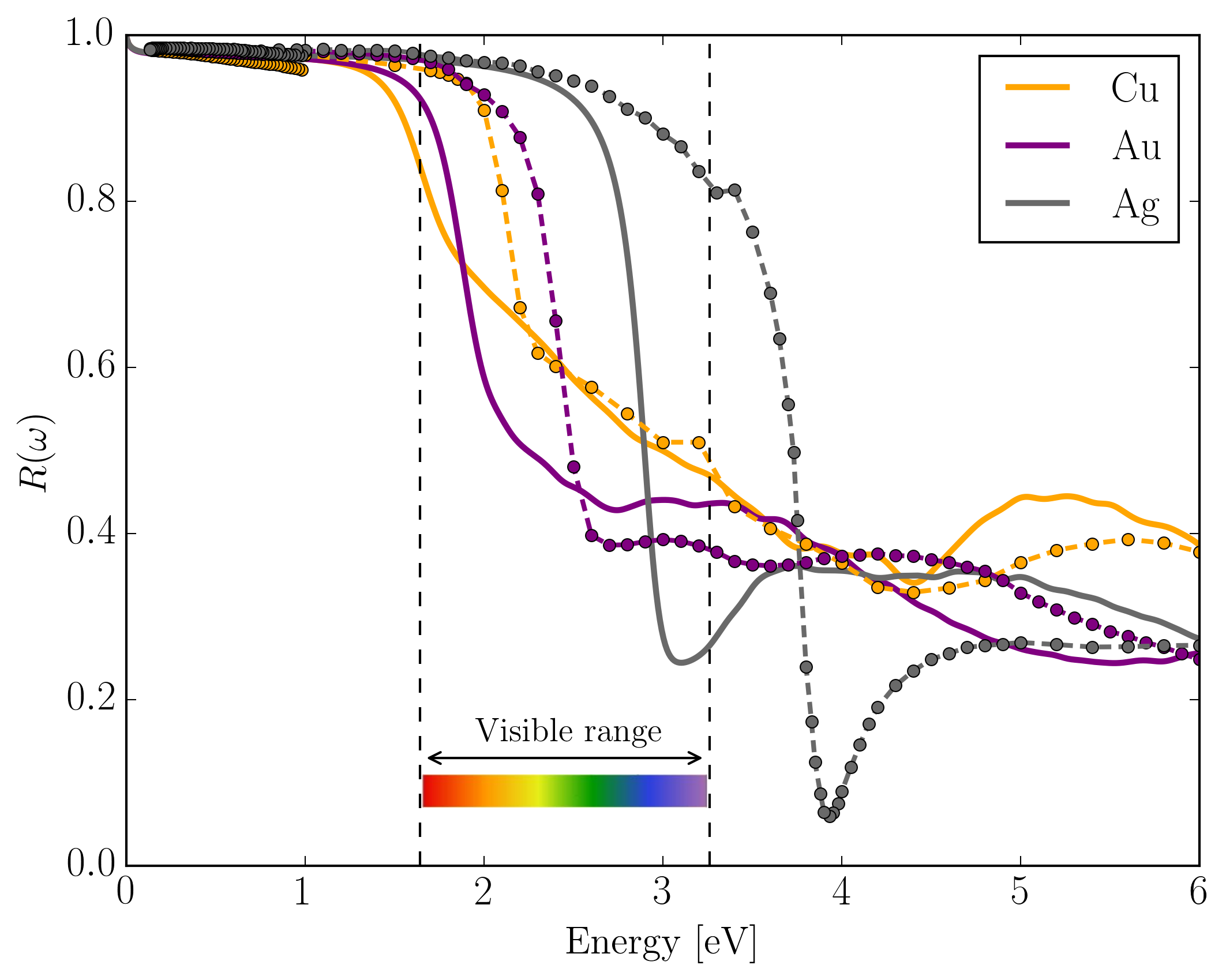}}

\subfloat{\includegraphics[scale=0.37]{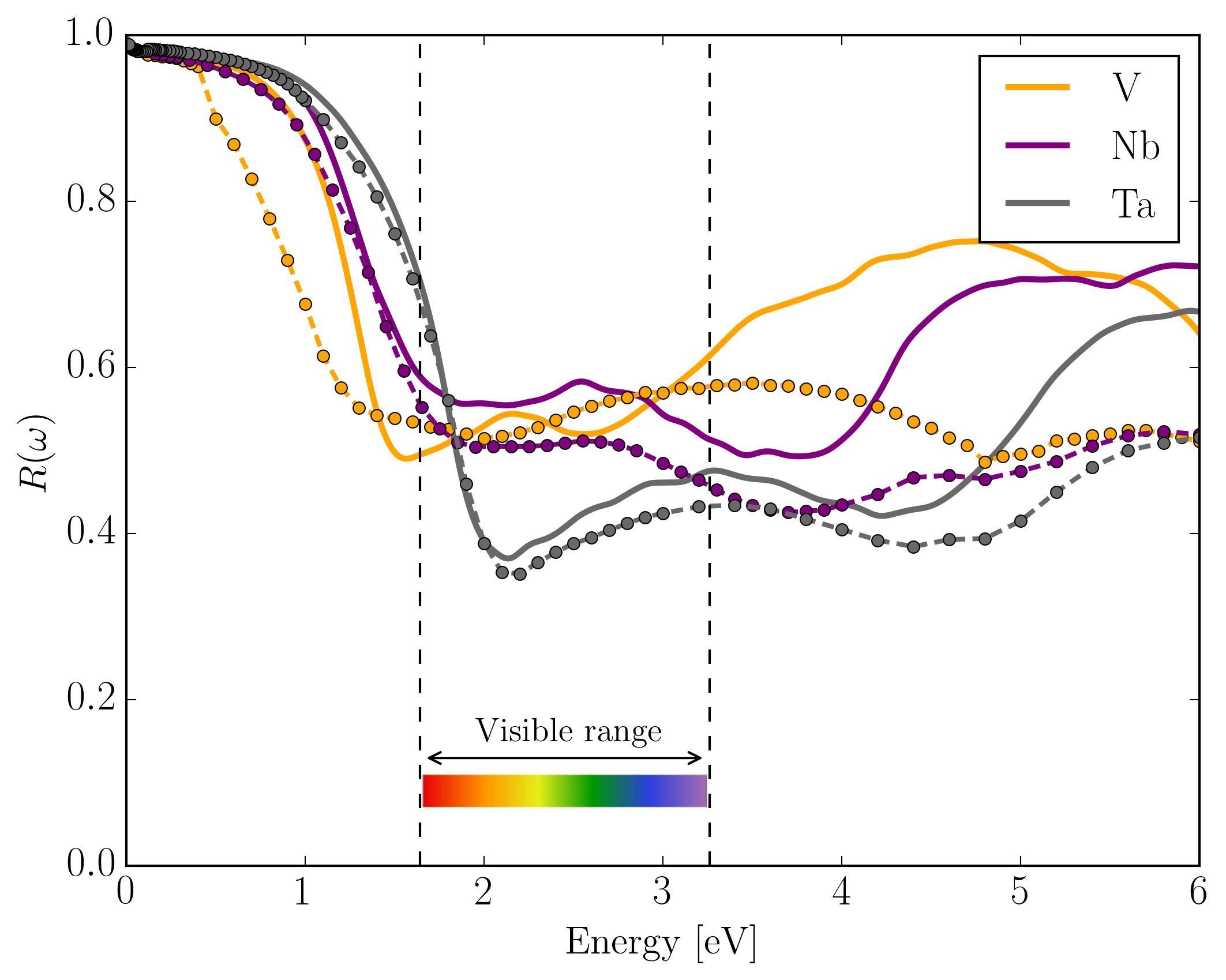}}
\subfloat{\includegraphics[scale=0.37]{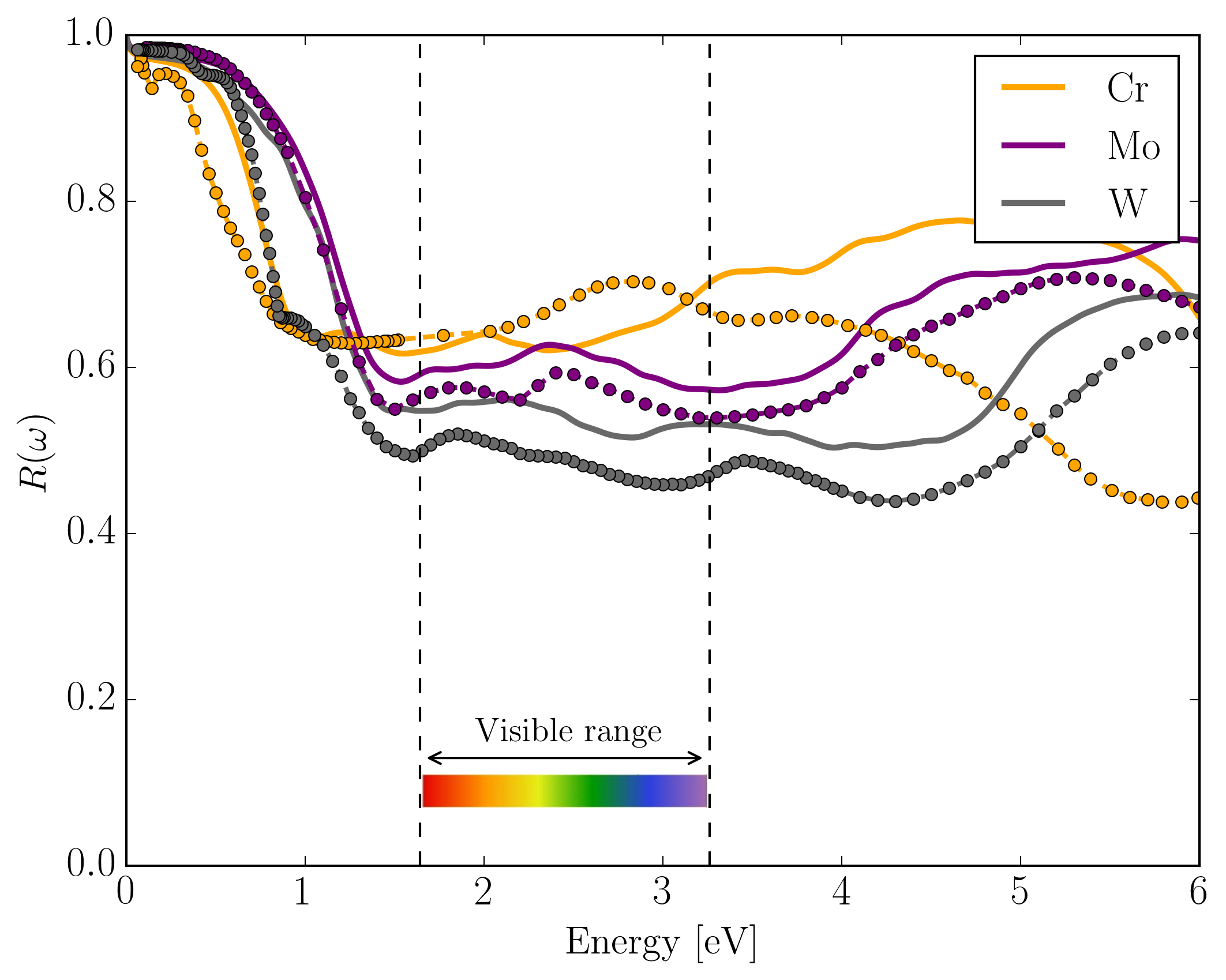}}

\caption{Simulated (solid lines) and experimental (dot-dashed lines) reflectivities for 18 elemental metals. Experimental data are taken from Ref.~\cite{Palik1998}. The two vertical dashed lines show the limits of the visible range.}
\label{fig:refl_elemental-metals_IP-vs-Exp}
\end{figure}

\begin{figure}[!hbtp]
\centering
\subfloat{\includegraphics[scale=0.3]{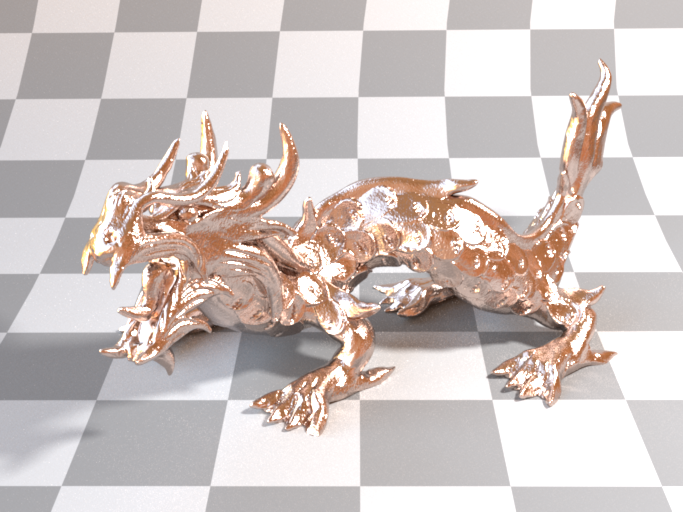}}
\subfloat{\includegraphics[scale=0.170]{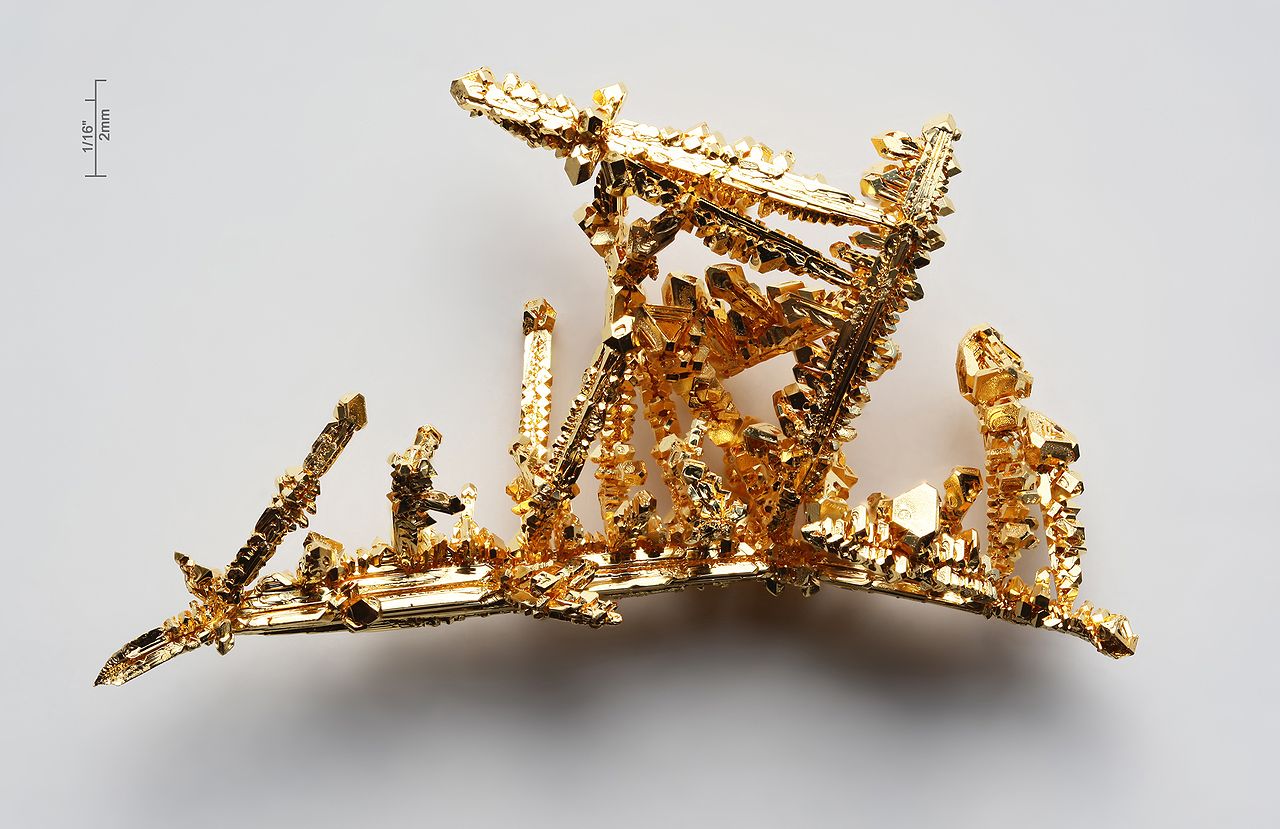}}

\subfloat{\includegraphics[scale=0.3]{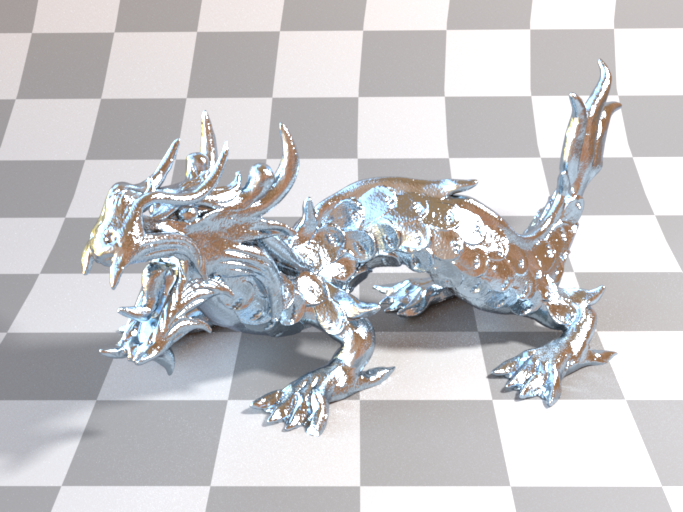}}
\subfloat{\includegraphics[scale=0.115]{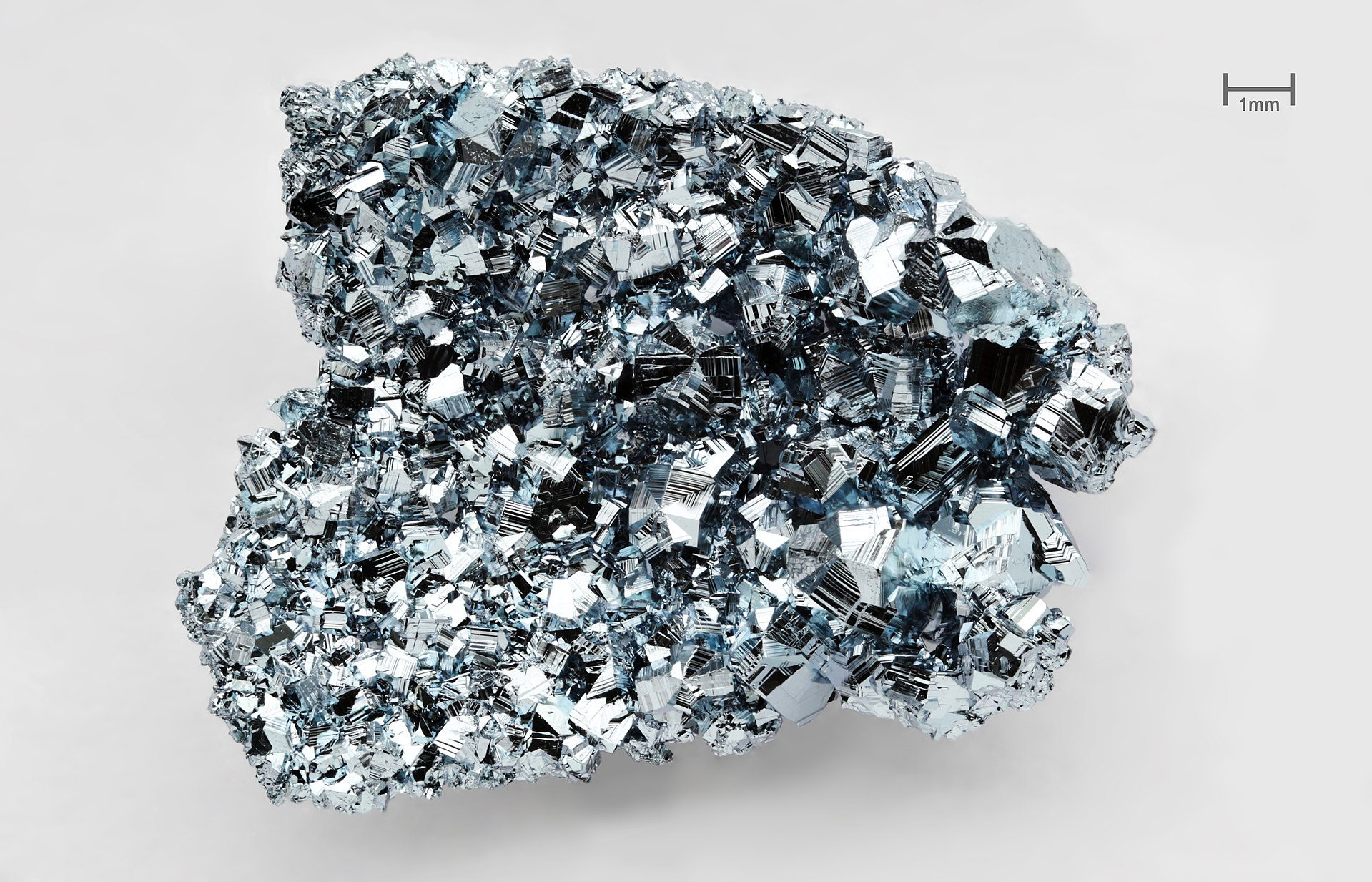}}

\subfloat{\includegraphics[scale=0.3]{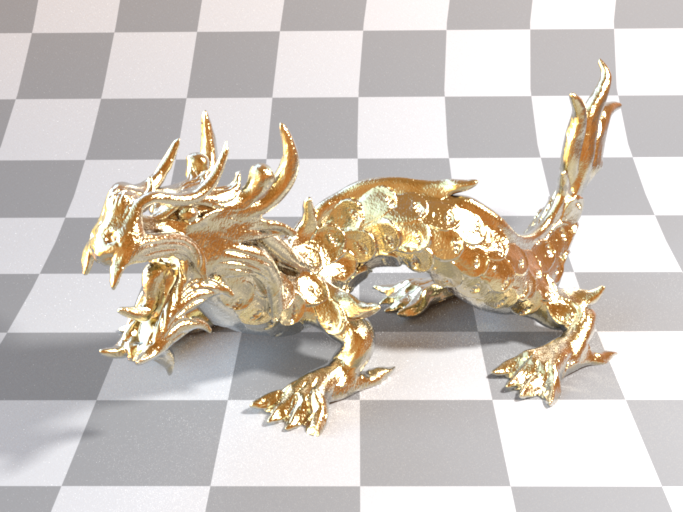}}
\quad \quad \,
\subfloat{\includegraphics[scale=2.75]{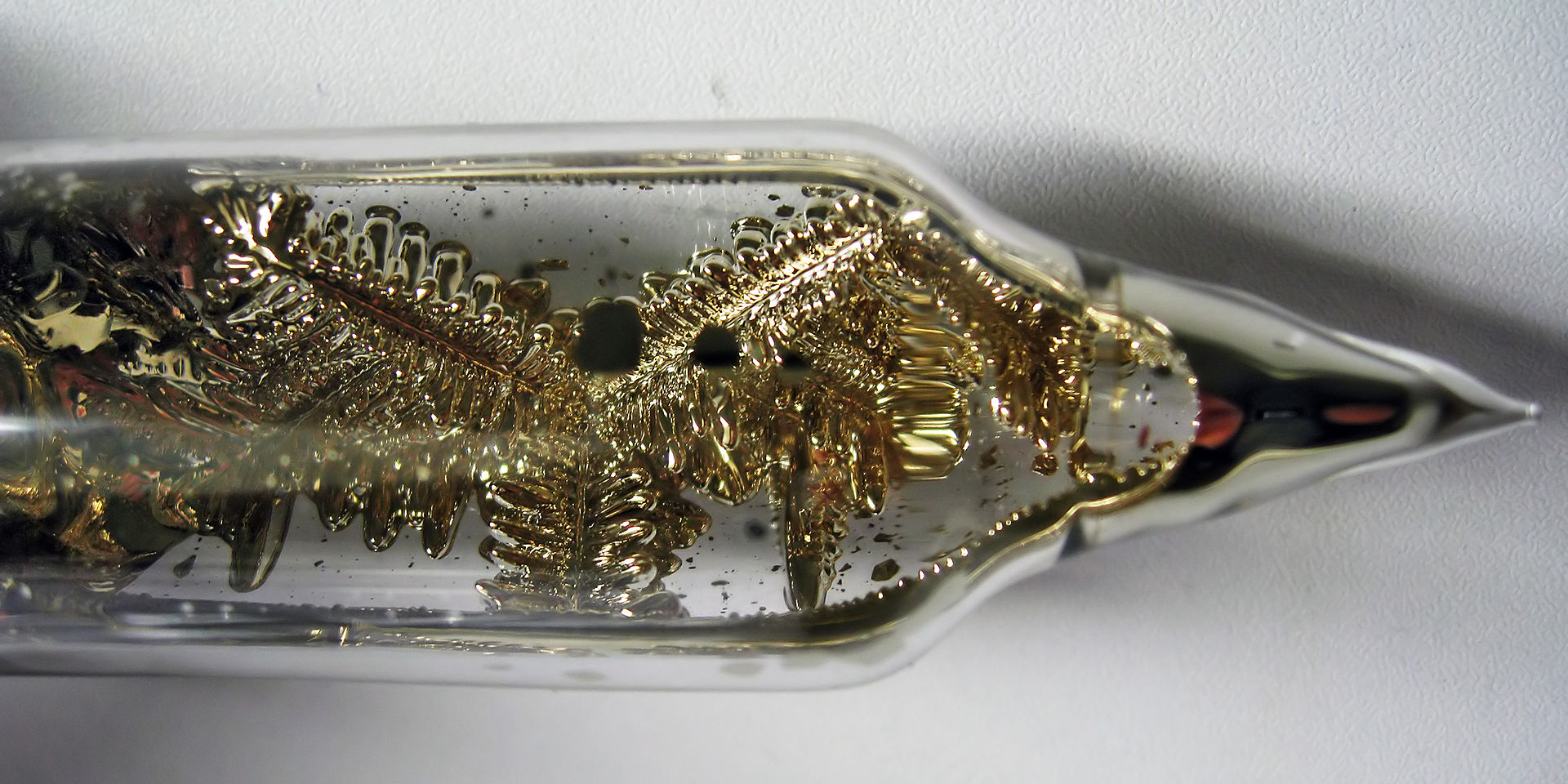}}
\caption{
Comparison between the simulated rendering of a metallic surface (left panel) and real samples (right panel) of pure gold (top), osmium (center) and caesium (bottom).
Photos of the gold and osmium samples are reproduced with the courtesy of Heinrich Pniok (www-pse-mendelejew.de) and are published under the Free Art License (http://artlibre.org/licence/lal/en/). Photo of the caesium sample is from the Dennis ``S.K." collection and is published under the GNU Free Documentation License (https://www.gnu.org/licenses/fdl.html).
}
\label{fig:rendering_elemental-metals}
\end{figure}

\noindent
Fig.~\ref{fig:refl_elemental-metals_IP-vs-Exp} shows the comparison between IPA results and experimental data for the reflectivity curves of 18 elemental metals, focusing on frequencies centered around the visible range (i.e. in the range [1.59, 3.26] eV). 
Experimentally, we observe high and flat reflectivities along the visible spectrum for the ``precious" transition metals (i.e. Rh, Ir and Pd) while we observe flat but slightly lower reflectivities for the other transition metals considered (i.e. V, Nb, Ta, Cr, Mo and W). 
As a consequence, in terms of CIELAB colour coordinates, metals in the first group have a large CIELAB brightness $L^*$ and thus whitish colour, while the others have smaller brightness and thus a more greyish colour (e.g. rhodium has $L^*$=90 while vanadium has $L^*$=78).  
An interesting exception among the transition metals is osmium, that shows a reflectivity curve that is low in the low-energy part of the visible spectrum but then suddenly rises in the blue-violet part, thus giving a bluish tint to pure osmium. A similar behaviour is found also in tantalum, but the rise of the reflectivity curve in the blue-violet region is significantly smaller and, consequently, also the bluish tint of the material is less pronounced. 
Instead, the simple $sp$ metals lithium, potassium and aluminium all have very high and nearly flat reflectivity curves in the visible range (and therefore whitish colour) while in beryllium the intensity of the reflectivity is lower, and comparable to that of the transition metals (and thus having a greyish colour). Interestingly, the reflectivity curve of caesium decreases significantly within the visible range, so that red and yellow radiation is strongly reflected, while all other visible frequencies are absorbed, giving a yellow tint to the material.
As clearly shown in Fig.~\ref{fig:refl_elemental-metals_IP-vs-Exp}, the IPA simulations reproduce  these different features of the elemental metals.
In contrast, for noble metals, while the characteristic drop in the reflectivity curve in the visible range (for Cu and Au) or in the ultraviolet (for Ag) is also reproduced by the simulations, it happens at smaller energies compared to experiments due to the well-known underestimation of the interband gap between valence $d$ bands and conduction $sp$ bands of PBE band structures.  
This discrepancy can be corrected using approaches beyond DFT, such as the $GW$ approximation of many-body perturbation theory. By correcting the DFT band energies at the $G_0W_0$ level, a quantitative agreement with respect to experiments is obtained for the optical spectra of Cu~\cite{Marini2001a} and Ag~\cite{Marini2002} but not for Au, for which $G_0W_0$ gives very similar results to PBE~\cite{Rangel2012}. For this latter case, the quasi-particle self-consistent $GW$ ($QSGW$)~\cite{vanSchilfgaarde2006, Kotani2007} approach is required for the occupied $5d$ bands of gold to be lowered in energy by the right amount~\cite{Rangel2012}. \\
A quantitative measure of the accuracy of the simulations can be obtained through the colour difference $\Delta E$ (given in Eq.~\ref{eq: delta_e}) with respect to experiments. Its average value is found to be $<\Delta E> = 6.4$.
For a more qualitative visual comparison, Fig.~\ref{fig:rendering_elemental-metals} shows the simulated rendering of a metallic surface of elemental gold, osmium and caesium together with the appearance of experimental samples of the same materials.
In gold, the shift of the reflectivity edge in the simulations with respect to experiments makes the rendered colour more reddish than the true red-yellow colour of pure gold. 
On the other hand the bluish colour of osmium and the yellow colour of caesium are well reproduced by the IPA simulations. \\
Moreover, as shown in Table~\ref{tab:drude_exp-vs-ip}, the IPA results for the Drude plasma frequency are in good agreement both with experiments and with previous simulations~\cite{Harl2008} performed at the same level of theory for some elemental metals. \\
From all these results, we conclude that the IPA approach applied on top of PBE band structures predicts the reflectivity and colour of elemental metals surprisingly well. Although the colour is not always in quantitative agreement with experiments, the shape and the main features of the experimental reflectivity curves are reproduced in elemental metals.
These results are somewhat surprising because it is known that quasi-particle corrections modify significantly the PBE band structure in metals and the corrections are k-dependent~\cite{Marini2001a, Marini2002} (i.e. they do not act as a simple scissor operator). Nonetheless, these approximated simulations manage to capture the correct features of the optical constants. This can intuitively be understood by the fact that the dielectric function is given by the sum of all possible vertical transitions over all the BZ and small differences in the positions and features of the bands (like gradient and curvature) are averaged out in the spectra. 
In the special case of noble metals, the position of the occupied $d$ bands in PBE is not correct and, since there are no other allowed interband transitions in that energy range, the onset of interband optical absorption (i.e. $\varepsilon_2^{\text{inter}}(\omega)$) in PBE is also not at the correct position (similar to the case of semiconductors for which the PBE band gap is systematically underestimated~\cite{Onida2002}). 
On the other hand, the shape of $\varepsilon(\omega)$ for noble metals is reasonably well reproduced. 

\subsection{Alloys}
\noindent
In order to validate the theoretical approach used on binary compounds, we first compare the reflectivity and colour between simulations and experiments for known coloured intermetallic compounds, as previously done for elemental metals. 
Second, we check the predictive accuracy of the simulations by studying in noble-metal-based alloys both the trends in reflectivity with respect to composition (in Ag-Au and Ag-Cu) and the differences in optical properties among different types of compounds for a given alloy composition (in Au-Cu and Ag-Cu).

\begin{figure}[!hbtp]
\centering
\includegraphics[width=\textwidth]{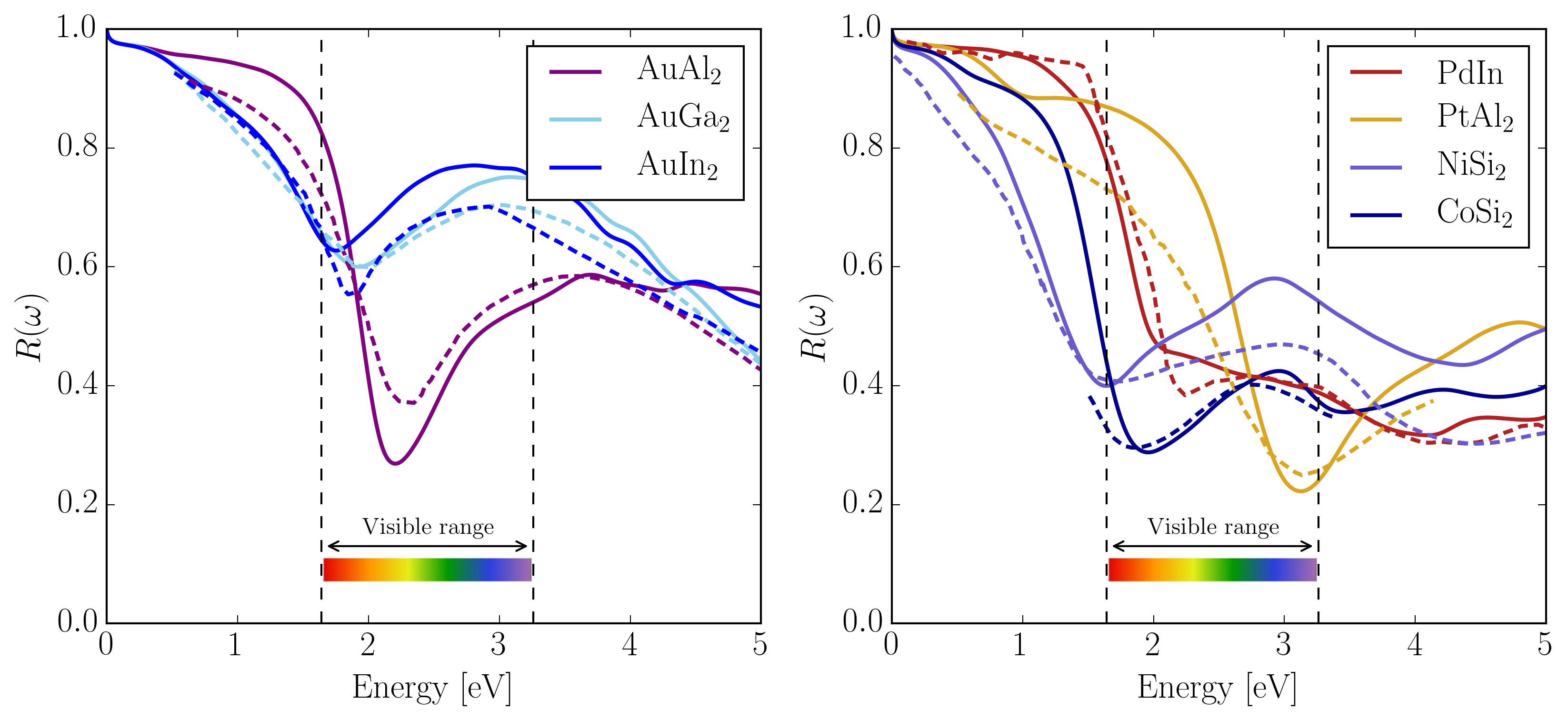}
\caption{Simulated (solid lines) and experimental (dashed lines) reflectivities of coloured intermetallics. Experimental data are taken from Ref.~\cite{Vishnubhatla1967} for AuAl$_2$, AuGa$_2$ and AuIn$_2$, from Ref.~\cite{Keast2013} for PtAl$_2$, from Ref.~\cite{Amiotti1990} for NiSi$_2$ and from Ref.~\cite{Steinemann1997} for CoSi$_2$ and PdIn. The two vertical dashed lines show the limits of the visible range.}
\label{fig:refl_intermetallic_sim-vs-exp}
\end{figure}

\begin{figure}[!hbtp]
\centering
\subfloat{\includegraphics[scale=0.3]{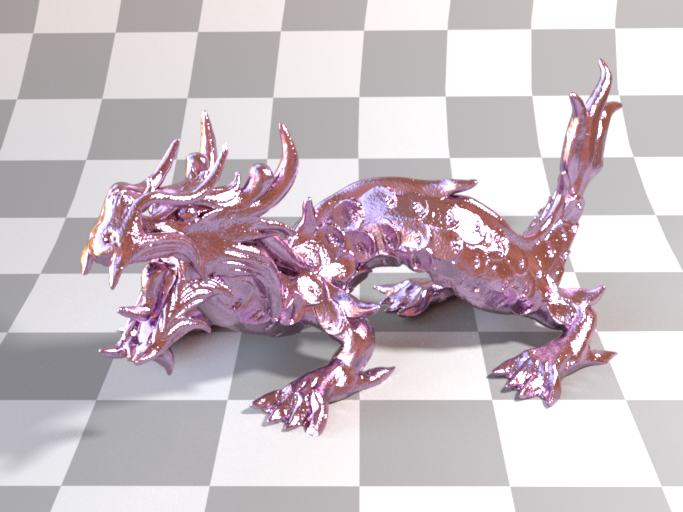}}
\subfloat{\includegraphics[scale=1.80]{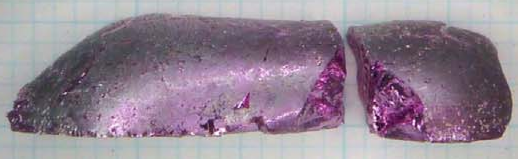}}

\subfloat{\includegraphics[scale=0.3]{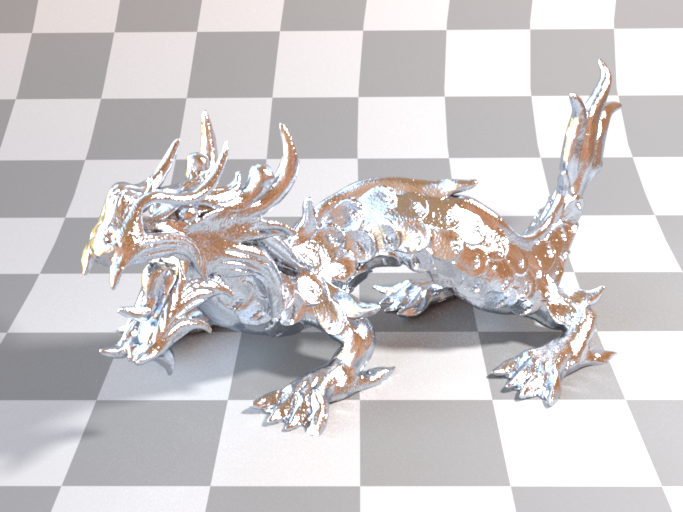}}
\subfloat{\includegraphics[scale=1.68]{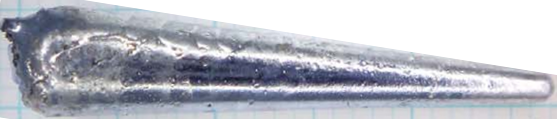}}

\subfloat{\includegraphics[scale=0.3]{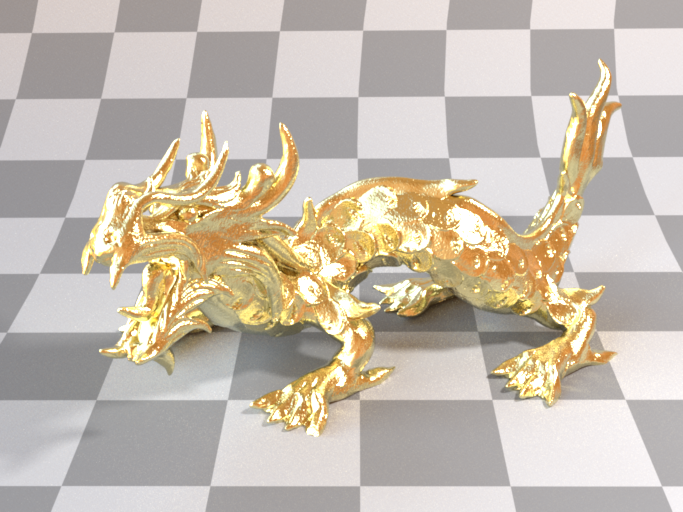}}
\quad \quad \quad  \qquad  \,\,\,\,
\subfloat{\includegraphics[scale=2.01]{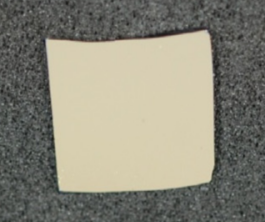}}
\caption{
Comparison between the simulated rendering of a metallic surface (left panel) and real samples (right panel) of the intermetallic compounds AuAl$_2$ (top), AuGa$_2$ (center) and PtAl$_2$ (bottom).
Images of the AuAl$_2$ and AuGa$_2$ samples are adapted from Ref.~\cite{Nishimura2018} while image of the PtAl$_2$ sample is adapted from Ref.~\cite{Furrer2014}.
}
\label{fig:rendering_intermetallic}
\end{figure}

\paragraph{Intermetallics.}
We first simulate the reflectivity and colour of intermetallic compounds that are experimentally known to be coloured.
The compounds studied are the purple AuAl$_2$, blue AuIn$_2$, bluish AuGa$_2$, yellow PtAl$_2$, red PdIn, blue-grey NiSi$_2$ and dark blue CoSi$_2$~\cite{Steinemann1997, Cretu1999}. 
All these intermetallics have cubic symmetry: AuAl$_2$, AuGa$_2$, AuIn$_2$, PtAl$_2$, CoSi$_2$ and NiSi$_2$ crystallize in the FCC CaF$_2$  prototype structure (space group Fm$\bar{3}$m) while PdIn crystallizes in the BCC CsCl prototype structure (space group Pm$\bar{3}$m). \\
As shown in Fig.~\ref{fig:refl_intermetallic_sim-vs-exp}, the experimental shape of the reflectivity curve for the coloured intermetallics is well reproduced by the simulations. 
The colour differences between simulations and experiments are summarized in Table~\ref{tab:colour_coloured-intermetallics_exp-vs-ip}, where the comparison with other first-principles simulations~\cite{Keast2011, Keast2013} is also reported.
The agreement with previous simulations is satisfactory and, moreover, we reproduce the true colour of the intermetallic compounds studied (although the CIELAB brightness is typically overestimated by the simulations). 
For example, the comparison between photorealistic rendering and real material samples clearly shows that the simulations predict the correct colours of purple AuAl$_2$, bluish AuGa$_2$ and yellow PtAl$_2$ (see Fig.~\ref{fig:rendering_intermetallic}). 
The characteristic colours of these highly symmetric intermetallic compounds are due to selective optical absorption in confined regions of the visible spectrum~\cite{Steinemann1997}.
For the gold compounds, the optical absorption inside the visible range is given by transitions from $sp$ conduction states below the Fermi level to unoccupied states above the Fermi level. The bands originating from the $5d$ states of gold, that are problematic in the study of elemental gold, are located at $\sim 5$ eV below the Fermi level and these do not contribute to the characteristic colours of these compounds~\cite{Keast2015}. This explains the better agreement with experiments found for the gold intermetallics compounds compared to the case of elemental gold.

\paragraph{Au-Ag-Cu.}

\begin{figure}[!hbtp]
\centering
\includegraphics[width=\textwidth]{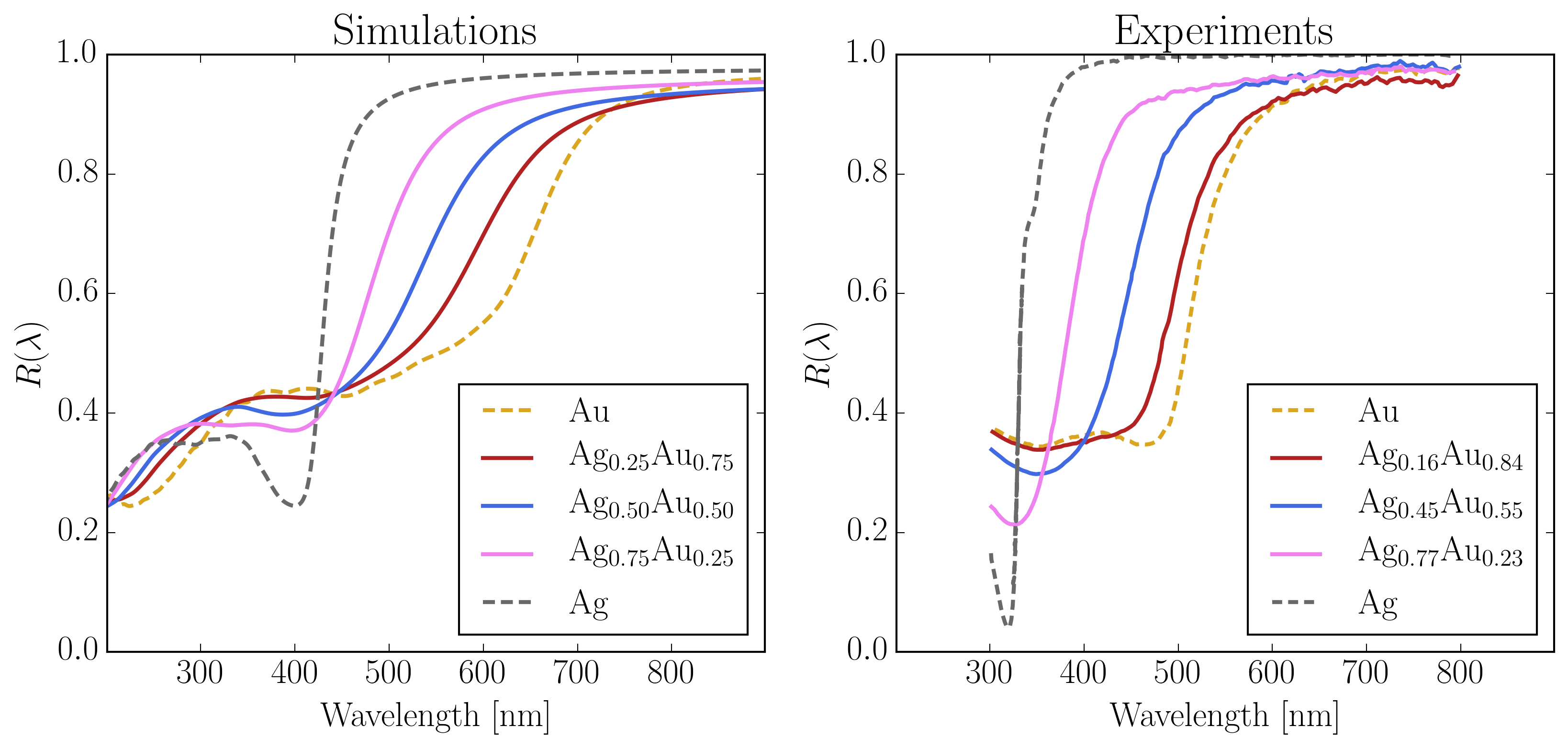}
\caption{Comparison of the trends in composition of the reflectivity inside the visible spectrum for Ag-Au solid solutions between SQS simulations (left panel) and experiments~\cite{Furrer2013} (right panel). For reference, we also report the reflectivity curves of elemental Au and elemental Ag (dashed lines). To note that the alloy compositions of experiments and simulations are not exactly the same.}
\label{fig:Ag-Au_refl_ip-vs-exp}
\end{figure}

\begin{figure}[!hbtp]
\centering
\includegraphics[width=\textwidth]{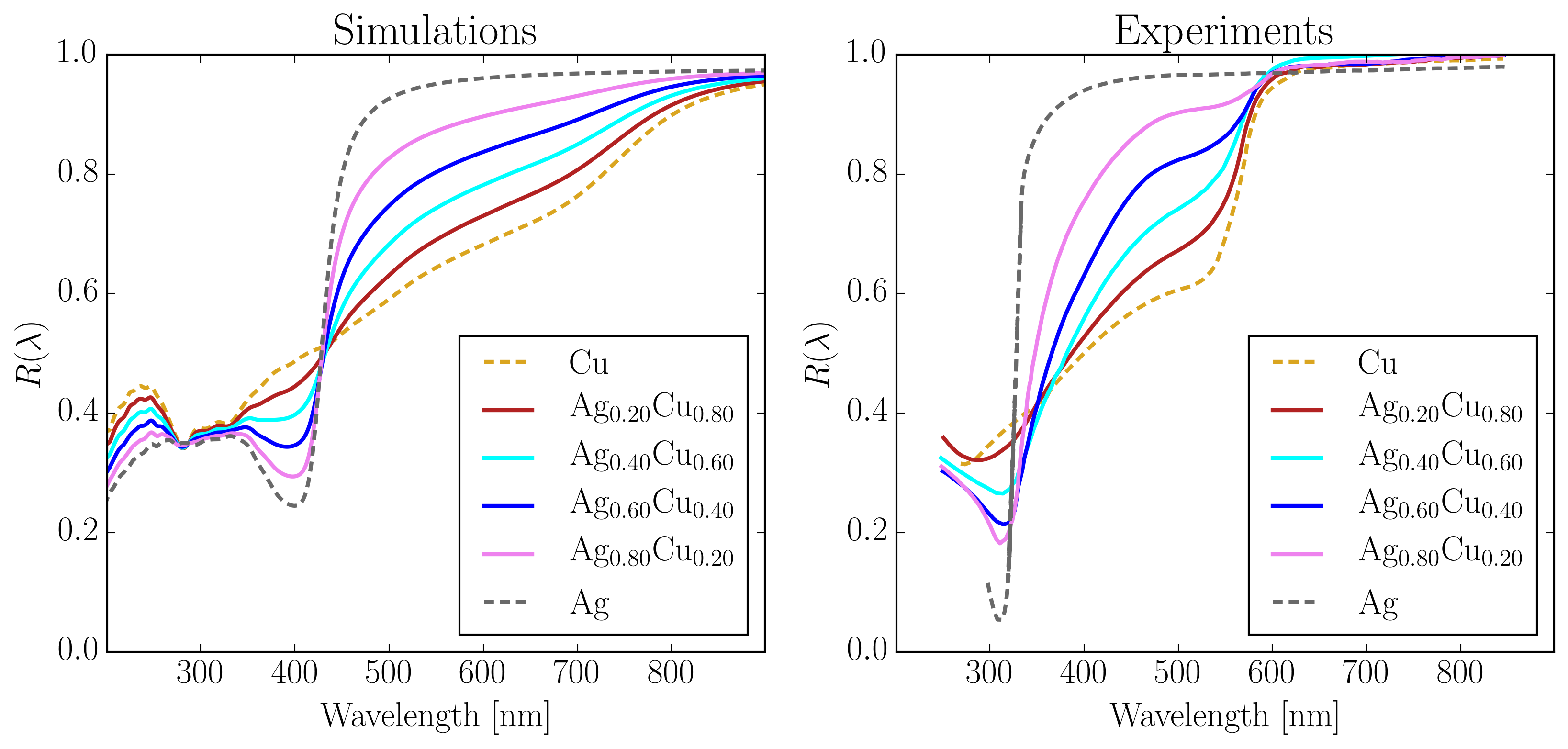}
\caption{Comparison of the trends in composition of the reflectivity inside the visible spectrum for Ag-Cu two-phase alloys between simulations (left panel) and experiments~\cite{Roberts1979} (right panel). The results of the simulations are obtained using the Bruggeman model in which the two phases of the system are assumed to be elemental Ag and elemental Cu (dashed lines).}
\label{fig:Ag-Cu_refl_ip-vs-exp}
\end{figure}

\begin{figure}[!hbtp]
\centering
\includegraphics[width=\textwidth]{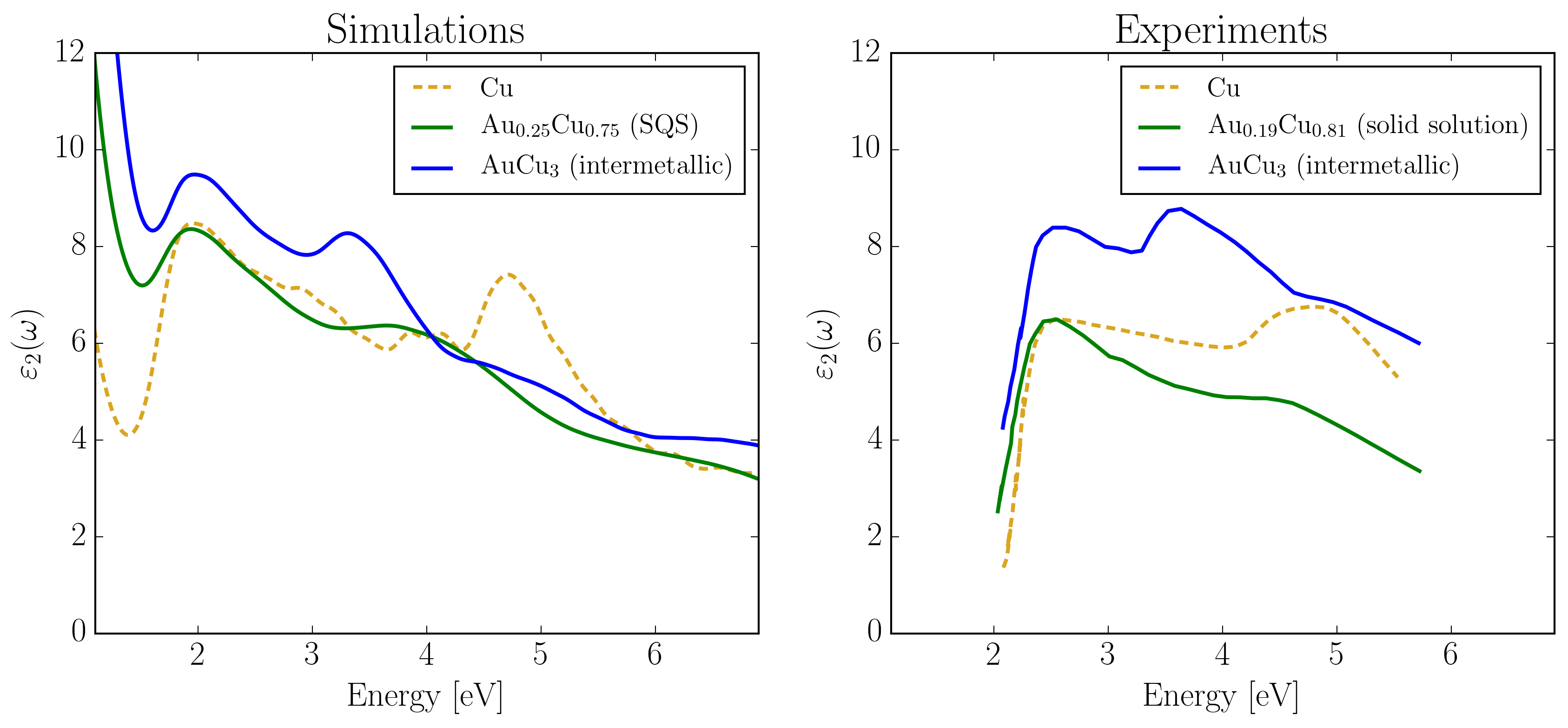}
\caption{Comparison of $\varepsilon_2(\omega)$ for Au$_{1-x}$Cu$_{x}$ between simulations (left panel) and experiments~\cite{Rivory1977} (right panel) for both the solid solution and the intermetallic phase AuCu$_3$ (at $x=0.75$ in the simulations and at $x=0.81$ in the experiments). For reference, we also report $\varepsilon_2(\omega)$ of elemental Cu (dashed lines). Experimental and simulated curves have been arbitrarily shifted along the vertical axis for clarity in the comparison.}
\label{fig:AuCu3_eps_ip-vs-exp}
\end{figure}

\begin{figure}[!hbtp]
\centering
\includegraphics[width=\textwidth]{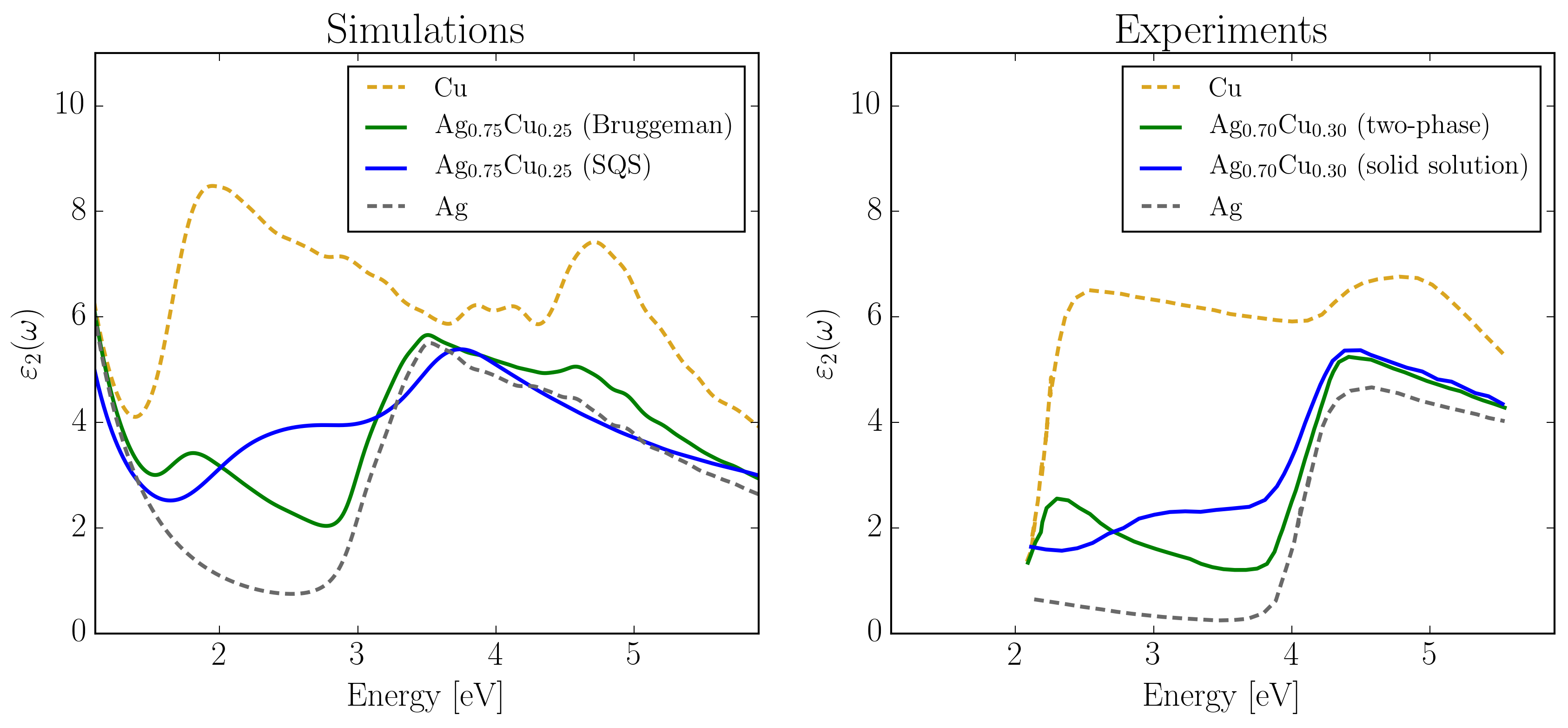}
\caption{Comparison of $\varepsilon_2(\omega)$ for Ag$_{1-x}$Cu$_{x}$ between simulations (left panel) and experiments~\cite{Rivory1977} (right panel) for both the solid solution and the two-phase alloy (at $x=0.25$ in the simulations and at $x=0.30$ in the experiments).}
\label{fig:Ag-Cu_eps_ip-vs-exp}
\end{figure}

\noi
The Au-Ag-Cu system is an ideal test case for the application of the computational approach described above to alloys since (i) several experimental optical data on this system are available, (ii) its constituent binaries show very different behaviours in terms of phase stability and so different types of compounds are observed and (iii) it is the basis of the most common jewellery and dental alloys in use today.  
Concerning the phase stability of the constituent binaries, Ag is completely soluble in Au thus Au and Ag form solid solutions for each composition and  no long-range order is observed at low temperatures.
Also Au and Cu form solid solutions over all concentrations at high temperatures but, for certain composition ranges, ordered intermetallic phases can be obtained at lower temperatures. In particular the known intermetallic compounds are the cubic AuCu$_3$ and Au$_3$Cu (space group Pm$\bar{3}$m), the low-temperature phase AuCu(I) (space group P4/mmm) and the high-temperature phase AuCu(II) (space group Imma). 
The phase diagram of Ag-Cu instead exhibits eutectic behaviour with a wide miscibility gap and the system tends to segregate in phases of nearly pure Ag and pure Cu at room temperature~\cite{Okamoto1990}. \\
We study the effect of composition on the reflectivity of the Ag-Au system and compare experimental data of solid solutions with SQS simulations.
Fig.~\ref{fig:Ag-Au_refl_ip-vs-exp} shows that the gradual shift to lower wavelengths of the reflectivity edge of gold by increasing the Ag content is reproduced by the simulations.
However, as already discussed above for the case of elemental noble metals, the position of the reflectivity edge in IPA simulations based on PBE band structures does not correspond to the experimental one, but it is instead systematically shifted to longer wavelengths for each atomic concentration $x$ considered. 
Although the simulations are not in quantitative agreement with experiments, the qualitative trends in reflectivity, and thus in colour, with respect to the alloy composition of Ag-Au are reproduced.
Similarly, we simulate the optical properties of Ag$_{1-x}$Cu$_x$ two-phase alloys by employing the Bruggeman model described above and study also in this system the effect of composition on the reflectivity.
The $\alpha$ and $\beta$ phases entering in the expression for the alloy dielectric function $\varepsilon_{\text{Br}}(\omega)$ of Eq.~\ref{eq: bruggeman_model} are assumed to be elemental Ag and elemental Cu, respectively. 
And the dielectric functions of the two constituent elements are taken from the simulations of elemental metals discussed above. 
As shown in Fig.~\ref{fig:Ag-Cu_refl_ip-vs-exp}, Ag additions in Cu increase the reflectivity at wavelengths shorter than the reflectivity edge of elemental Cu but do not shift, as it happens in Ag-Au solid solutions, the position of the edge. 
The Bruggeman model provides the correct trend with composition but the effect on the drop in the reflectivity is less evident because the reflectivity edge of elemental Cu in IPA simulations is less steep than the experimental one. Note that the application of the Bruggeman model to experimental data of the dielectric function of elemental Ag and elemental Cu gives very good agreement with experimental data for the two-phase alloy and validates the use of the model. \\
Summarizing, for Ag-Au solid solutions, where there is a gradual shift of the reflectivity edge by varying alloying additions from elemental Au to elemental Ag, the colour of the alloy changes from red-yellow to yellow, pale greenish-yellow and eventually white of pure Ag.  Au-Cu solid solutions show a similar behaviour~\cite{Rivory1977} and the colour of the alloy changes from red-yellow to reddish and eventually red of pure Cu.
Instead, in Ag-Cu two-phase alloys there is no shift of the reflectivity edge but, for all wavelengths in the visible range below the reflectivity edge of elemental Cu, the reflectivity curve rises roughly uniformly so that the colour of Ag-Cu changes from the red of pure Cu to reddish and then directly to whitish and white of pure Ag~\cite{Cretu1999}. \\
After considering the effect of composition on the reflectivity of binary alloys, we now study the effect of different types of compounds for a given fixed atomic concentration $x$ directly on the dielectric function of the Au-Cu and Ag-Cu systems. 
Indeed, for Au-Cu at the composition $x=0.81$, experimental data are available in the literature for the optical absorption of both the solid solution and the intermetallic compound AuCu$_3$~\cite{Rivory1977}.  
Analogously, for Ag-Cu at the composition $x=0.30$, experimental data are available for both a segregated two-phase sample made of a pure Cu phase and a pure Ag phase, and for a metastable solid solution obtained by vapor quenching~\cite{Rivory1977}.
We compare the optical absorption of the Au$_{1-x}$Cu$_x$ solid solution, at $x = 0.81$ in experiments and at $x=0.75$ in simulations, with the optical absorption of the intermetallic compound appearing around the composition $x=0.75$, i.e. the cubic AuCu$_3$ phase.
The purpose of this comparison is to study the differences in optical properties between ordered and disordered phases.
As shown in Fig.~\ref{fig:AuCu3_eps_ip-vs-exp}, the optical absorption of the intermetallic compound is very similar to the one of the random alloy with the notable exception of the presence of an additional peak at around 3.6 eV, which is missing in $\varepsilon_2(\omega)$ for the solid solution. 
The comparison of the SQS results for the disordered alloy with the simulated results of the intermetallic compound shows that the simulations clearly capture this small difference. 
Nonetheless, we underline that there is no significant change in the resulting colour between ordered and disordered alloy for this system because the position of the onset of optical absorption is not modified by the presence of long-range order, and thus neither is the colour.\\
Similarly, Fig.~\ref{fig:Ag-Cu_eps_ip-vs-exp} reports the comparison between the optical absorption of the Ag$_{1-x}$Cu$_x$ two-phase alloy, at $x = 0.70$ in experiments and at $x=0.75$ in simulations, with respect to that of the metastable solid solution having the same composition.
In the two-phase alloy, where the alloy optical properties are well approximated by a combination of those of pure Cu and pure Ag (Bruggeman model), we observe two onsets of absorption: the first one at $\sim$ 2.1 eV corresponding to the absorption edge of pure Cu and the second one at $\sim$ 4.0 eV corresponding to the absorption edge of pure Ag.
The optical absorption of the solid solution instead is very similar to the one of pure Ag but, in addition, we observe the presence of a supplementary broad peak at energies below the onset of absorption of pure Ag due to Cu impurity states.
The SQS results for the solid solution and the results of the Bruggeman model applied on the IPA dielectric function of elemental Ag and Cu reproduce the two different trends, although the SQS shows a small blueshift of the peak that follows the absorption edge of pure Ag which is not observed experimentally.

\section{Discussion}
\noindent
We have shown that the theoretical methods and approximations considered in this paper, i.e. IPA optical spectra computed on top of the DFT-PBE electronic structure, can be employed in systematic studies on the optical properties of metals in order to predict trends in real metallic systems and to help the search for novel materials with specific optical properties, and therefore also colours, by exploring the composition space through the computational screening of materials~\cite{Prandini2019}.
Moreover, this work could help stimulate future studies aiming to achieve the photorealistic simulation of different types of materials by means of first-principles techniques. 
For example, the systematic validation of the approach performed on elemental metals and binary alloys can be seen as a necessary preliminary step for the photorealistic simulation of more complex metallic alloys having a larger number of constituent elements, such as ternaries, quaternaries, etc., which are more relevant for technological applications (e.g. superalloys and high-entropy alloys). 


\section{Methods}
\noindent

\noi
\subsection{Workflow}
\noi
All DFT calculations are performed with the Quantum ESPRESSO distribution~\cite{Giannozzi2009}, which is based on the plane-wave pseudopotential method for the numerical solution of the KS equations.
We use Shirley's interpolation method~\cite{Shirley1996, Prendergast2009} as implemented in the \texttt{SIMPLE} code~\cite{simple_cpc} to evaluate the IPA dielectric function of metals including both interband and intraband contributions.
Photorealistic rendering is performed with the Mitsuba renderer~\cite{mitsuba_webpage}.
Pseudopotentials and plane-wave cutoffs are chosen according to the results of the standard solid-state pseudopotential (SSSP) protocol~\cite{sssp_npj} in order to have reliable and converged band structures as the starting ingredients for the evaluation of the IPA dielectric function. Since the \texttt{SIMPLE} code supports only norm-conserving pseudopotentials, we use optimized norm-conserving Vanderbilt (ONCV)~\cite{Hamann2013} pseudopotentials from the SG15~\cite{Schlipf2015} and PseudoDojo~\cite{Dojo2017} PBE pseudopotential libraries for all elements considered (see Supplementary Discussion 2 for more details on the choice of the pseudopotentials from the SSSP database of tests).
For the purpose of automation, the sequence of calculations required by the computational approach described in this work is implemented as a workflow within the framework of the AiiDA~\cite{Pizzi2016} infrastructure for computational science. Thanks to this ColourWorkflow (see Fig.~\ref{fig:colourworkflow}), it is possible, giving as input a generic crystal structure, to obtain directly as output the reflectivity and colour of a given material. \\
In all simulations, relativistic effects are accounted for at the scalar-relativistic level (see Supplementary Discussion 3 for an analysis on the effect of spin-orbit coupling on the optical properties of heavy elements) while the IPA dielectric function is always evaluated by including the non-local contribution of the pseudopotentials in the computation of the velocity matrix elements, as implemented in \texttt{SIMPLE}.

\subsection{Elemental metals}
\noi
All calculations on elemental metals are performed on the ground-state crystal structures at zero temperature, as provided in Ref.~\cite{Lejaeghere2014}. 
The equilibrium volume of each structure corresponds to the reference PBE value obtained by extensively tested all-electron calculations for the equation of state~\cite{Lejaeghere2016}.
If needed, the crystal structures are reduced to the primitive cell using the spglib library~\cite{spglib_webpage}. 
Spin-polarization is not included in our calculations.
In the self-consistent DFT calculations for the evaluation of the ground-state density we use a Monkhorst-Pack grid~\cite{MonkhorstPack} of $24 \times 24 \times 24$ and a cold smearing~\cite{Marzari1999} of 0.02 Ry. In the non self-consistent band structure calculations needed for the construction of the Shirley's basis we use a uniform k-grid of $2 \times 2 \times 2$ including the seven periodic images of the $\Gamma$-point of the BZ and at least 30 empty conduction bands.
From a convergence study on the dielectric function we decide to employ an interpolation k-grid of $64 \times 64 \times 64$ and $\eta=\gamma=0.1$ eV in \texttt{SIMPLE} for each elemental metal considered, with the exception of elemental aluminium for which, because of a very slow convergence of $\varepsilon^{\text{inter}}(\omega)$ with respect to k-points sampling, the interpolation k-grid used is $80 \times 80 \times 80$ and $\eta$ is set to 0.2 eV. The Shirley's basis is constructed setting the threshold for the Gram-Schmidt orthonormalization algorithm equal to 0.0075 a.u. (input variable named $s_b$ in \texttt{SIMPLE}).
 
\subsection{Alloys} 
\noi
For the simulation of all binary compounds considered, we always use as plane-wave cutoff the largest value between the plane-wave cutoffs of the two constituent elements, as taken from Supplementary Table 1. The Shirley's basis is constructed setting $s_b = 0.01$ a.u. in \texttt{SIMPLE} and considering a number of empty bands at least equal to the number of occupied bands.
We choose the interpolation k-grid to be used in the evaluation of the dielectric function in terms of a k-point density, which is defined as the maximum distance between adjacent k-points along the reciprocal axes (in \AA$^{-1}$).
For all the seven cubic intermetallic compounds considered we select as k-point density 0.04 \AA$^{-1}$. With this choice the number of k-points included in the uniform k-grids is of the order $O(10^{5})$, which correspond to uniform k-grids in the range from $46 \times 46 \times 46$ up to $56 \times 56 \times 56$. \\
All the SQSs used in this work to simulate solid solutions of the systems Ag$_{1-x}$Au$_x$, Au$_{1-x}$Cu$_x$ and Ag$_{1-x}$Cu$_x$ are generated with the ATAT package~\cite{vandewalle2002, vandewalle2013}.
Since we consider only the simple stoichiometric ratios $x=0.25,0.5,0.75$, we use small FCC SQSs with 16 atoms per cell. 
The interpolation k-grid is set according to a k-point density of 0.04 \AA$^{-1}$ (corresponding roughly to 11'000 points in the BZ). 


\section*{Acknowledgements}
\noindent
The authors warmly thank Fanny Lalire and Fr\'ed\'eric Diologent for several useful discussions and for sharing with us confidential experimental results.

\section*{Competing Interests}
\noindent
The authors declare no competing interests.

\section*{Author contributions}
\noindent
N. M. and G.-M. R. designed the study; G. P. developed the computational workflow, performed the calculations and wrote the manuscript. All authors discussed and analysed the results and commented on the manuscript.

\section*{Funding}
\noindent
This research was supported by Varinor SA (CH 2800 Del\'emont, Switzerland).

\section*{Data availability}
\noindent
The data that support the findings of this study are available from the corresponding
authors upon reasonable request. 
The source code of the ColourWorkflow and the input scripts necessary in order to reproduce the simulations performed for this work are available at https://github.com/giprandini/colour-workflow.

\section*{Additional information}
\noindent
\textbf{Supplementary information} is available at \textit{npj Computational Materials} website.\\


\section{References}
\bibliographystyle{naturemag}

\clearpage
\newpage

\section{Figures}

\clearpage
\newpage

\section{Tables}

\begin{table}[!hbtp]
\centering
\renewcommand{\tabcolsep}{5mm}
\begin{tabular}{ll}
\hline
\hline
Binary alloys & Computational method \\
\hline
Intermetallic compounds  &  Primitive cell \\
Solid solutions  &  Supercell (SQS) \\
Heterogeneous alloys  &  Bruggeman model \\
\hline
\end{tabular}
\caption{Different types of compounds and corresponding simulation method used in this work for the first-principles simulation of these systems.}
\label{tab:alloy-type_simulations}
\end{table}

\begin{table}[!hbtp]
\centering
\renewcommand{\tabcolsep}{5mm}
\begin{tabular}{llll}
\hline
\hline
Element & This work & J. Harl~\cite{Harl2008} &  Exp.  \\
\hline
Cu & 8.8 & 9.1 & 8.8, 8.9 \\
Ag & 8.9 & 9.2 & 8.9 $\pm$ 0.2 \cite{Yang2015}, 8.9 \\
Au & 8.6 & 9.0 & 8.45 \cite{Olmon2012}, 8.7 \\
Li & 6.4 & 6.5 & 6.4 \\
Na & 6.0 & 5.9 & 5.7 \\
Ca & 4.1 & 4.3 & 5.7 \\
Al & 12.5 & 12.6 & 12.3 , 12.5  \\
Rh & 9.6 & 10.1 & \\
Pd & 7.0 & 7.4  & \\
Pt & 8.4 & 8.8  & \\
\hline
\end{tabular}
\caption{Computed values of the IPA Drude plasma frequency $\omega_{\text{D}}$ (in eV) compared to previous simulations (J. Harl~\cite{Harl2008}) and experiments (Exp.). The experimental values with no explicit reference are extracted from the data reported in Ref.~\cite{Harl2008}. For transition metals there are no experimental data available because, due to the presence of interband transitions even at vanishingly small frequencies, the Drude plasma frequency cannot be extracted by fitting experimental optical data to the Drude model, even at very low energies.}
\label{tab:drude_exp-vs-ip}
\end{table}

\begin{table}[!hbtp]
\centering
\renewcommand{\tabcolsep}{5mm}
\begin{tabular}{lll}
\hline
\hline
Compound & $\Delta E_{\text{exp}}$ & $\Delta E_{\text{sim}}$ \\
\hline
AuAl$_2$  &  11~\cite{Vishnubhatla1967} & 8~\cite{Keast2013}, 4~\cite{Keast2011} \\
AuGa$_2$  &  2~\cite{Vishnubhatla1967}  & 1~\cite{Keast2011} \\
AuIn$_2$  &  4~\cite{Vishnubhatla1967}  & 1~\cite{Keast2011} \\
PtAl$_2$  &  12~\cite{Keast2013}  & 2~\cite{Keast2013} \\
CoSi$_2$  &  3~\cite{Steinemann1997}   &  \\
NiSi$_2$  &  5~\cite{Amiotti1990}  &  \\
PdIn      &  10~\cite{Steinemann1997}  &  \\
\hline
\end{tabular}
\caption{Colour differences in CIELAB space between simulated colours (present work) and experimental colours~\cite{Vishnubhatla1967, Keast2013, Steinemann1997, Amiotti1990} derived from reflectivity data, $\Delta E_{\text{exp}}$, and between simulated colours (present work) and previously published simulations~\cite{Keast2013, Keast2011}, $\Delta E_{\text{sim}}$.}
\label{tab:colour_coloured-intermetallics_exp-vs-ip}
\end{table}

\end{document}